%% file: paper.tex
\documentclass[fleqn,usenatbib]{mnras}
\usepackage{amsmath}
\usepackage{amssymb}

\usepackage{geometry}
\usepackage{refstyle}
\usepackage{units}
\usepackage{url}
\usepackage{varwidth}
\usepackage{graphicx}
\usepackage[authoryear]{natbib}

\makeatletter

\AtBeginDocument{\providecommand\Figref[1]{\ref{Fig:#1}}}
\AtBeginDocument{\providecommand\Tabref[1]{\ref{Tab:#1}}}
\AtBeginDocument{\providecommand\Subsecref[1]{\ref{Subsec:#1}}}
\providecommand\textquotedblplain{%
  \bgroup\addfontfeatures{RawFeature=-tlig}\char34\egroup}
\providecommand{\tabularnewline}{\\}
\newenvironment{cellvarwidth}[1][t]
    {\begin{varwidth}[#1]{\linewidth}}
    {\@finalstrut\@arstrutbox\end{varwidth}}
\RS@ifundefined{subsecref}
  {\newref{subsec}{name = \RSsectxt}}
  {}
\RS@ifundefined{thmref}
  {\def\RSthmtxt{theorem~}\newref{thm}{name = \RSthmtxt}}
  {}
\RS@ifundefined{lemref}
  {\def\RSlemtxt{lemma~}\newref{lem}{name = \RSlemtxt}}
  {}

\usepackage{newtxtext,newtxmath}

\usepackage[T1]{fontenc}

\pubyear{\the\year{}}

\usepackage{xcolor}
\definecolor{fuchsia}{RGB}{230,40,230}

\usepackage[titletoc]{appendix}
\usepackage{xspace}
\newcommand{\elasticc}{\textsc{ELAsTiCC}\xspace}
\date{PREPRINT ONLY, not yet submitted, PLEASE CHECK BACK FOR UPDATES}

\title{\nicknameWithEmoji: Astronomical Timeseries CAusal Transformer}
\author[zora tung]{
zora tung,$^{1}$
\\
$^{1}$Independent researcher.
}

\makeatother

\begin{document}

\newcommand{\nicknameWithEmoji}{ATCAT (@\raisebox{-0.57ex}{\includegraphics[height=2.57ex]{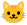}})\xspace}
\label{firstpage}
\maketitle

\input{results/macros.tex}

\begin{abstract}

The Legacy Survey of Space and Time (LSST) at the Vera C. Rubin Observatory
will capture light curves (LCs) for 10 billion sources and produce
millions of transient candidates per night, necessitating scalable,
accurate, and efficient classification. To prepare the community for
this scale of data, the Extended LSST Astronomical Time-Series Classification
Challenge (\elasticc) sought to simulate a diversity of LSST-like
time-domain events. Using a small transformer-based model and refined
light curve encoding logic, we present a new state of the art classification
performance on \elasticc, with \res{our lc_only f1}\% F1 on LC-only
classifications, and \res{our lc_meta f1}\% F1 on LC+metadata classifications.
Previous state of the art was 65.5 $\pm$ 0.28\% F1 for LC-only, and
for LC+metadata, 84\% F1 with a different setup and \res{atat lc_meta f1}\%
F1 with a directly comparable setup. Our model outperforms previous
state-of-the-art models for fine-grained early detection at all time
cutoffs, which should help prioritize candidate transients for follow-up
observations. We demonstrate label-efficient training by removing
labels from 90\% of the training data (chosen uniformly at random),
and compensate by leveraging regularization, bootstrap ensembling,
and unsupervised pretraining. Even with only 10\% of the labeled data,
we achieve \res{pretr setup A lc_only f1}\% F1 on LC-only and \res{pretr setup A lc_meta f1}\%
F1 on LC+metadata, validating an approach that should help mitigate
synthetic and observational data drift, and improve classification
on tasks with less labeled data. We find that our base model is poorly
calibrated via reliability diagrams, and correct it at a minimal cost
to overall performance, enabling selections by classification precision.
Finally, our GPU-optimized implementation is \res{our model times faster vs oracle}$\times$
faster than other state-of-the-art \elasticc models, and can run
inference at \textasciitilde\res{perfbench our model lc meta rtx 4090 rounded}
LCs/s on a consumer-grade RTX 4090 GPU, making it suitable for large-scale
applications. It is also correspondingly cheaper to train, making
it accessible to more researchers.

\end{abstract}

\section{Introduction}

The LSST at the Vera C. Rubin Observatory offers the potential to
uncover millions of transient events, both to improve our understanding
of known objects and to find novel ones \citep{lsstsciencebook2009}.
However, it is challenging to identify these transients, because most
new ones are faint, single-pixel sources. Additionally, the scale
of data requires the use of automated techniques \citep{Fraga_2024}.
Classification systems are used to distinguish interesting transient
objects from variable stars and less interesting ones, to allow for
immediate follow-up spectroscopy, further optical photometry, and
imaging in other wavebands. They are also used for population studies,
such as improving the mapping of the Milky Way by identifying more
RR Lyrae, and detecting a large number of supernovae to study the
dependence of dark energy properties on direction. Astronomers also
hope to find rare objects and events such as novae and stellar flares,
gamma-ray bursts and X-ray flashes, active galactic nuclei, stellar
disruptions by black holes, and evidence of neutron stars and black
hole binaries; some of these require timely follow-up, and others
exist for a longer time, but are difficult to identify (see \citealt{Ivezić_2008}
for an overview). The quality of classification systems determines
how much value can be obtained through timely follow-up observations,
and the quality of downstream scientific analyses. This motivates
our work.

Classification models for time-domain astronomy encounter several
challenges, which set them apart from traditional machine learning
(ML) models such as image classification or natural language processing.
In the \elasticc dataset, which is synthetic but carefully designed
to mimic real data, light curves not only have extreme values and
occasionally feature high observational noise, but are also irregularly
sampled, and each “channel” (band-pass color filter) is sampled
at a different time. For example, a given object may have a flux value
in the \texttt{u} band one day, and then an observation in the \texttt{g}
band 2 days later; this corresponds to physical color filter plates
being changed on the telescope. Objects may only be visible during
part of the year. In this manner, the problem may appear more in the
domain of traditional statistics, but well-tuned ML models tend to
perform better, likely reflecting the high-dimensional, complex nature
of the underlying objects being modeled.

\citet{Cabrera-Vives_2024} demonstrated the viability and superiority
of neural network approaches for \elasticc, by both creating a strong
random forest (RF) baseline model, and outperforming it with a 3-layer
transformer model (ATAT). Several notable classifiers exist in the
space of LSST classification; in 2018, the (Photometric LSST Astronomical
Time-Series Classification Challenge) (PLAsTiCC) Kaggle challenge
was run; resulting classification models include \citet{Boone_2019}
and \citet{Qu_2021}. The \elasticc dataset is largely seen as its
successor, and classification models for it include \citet{Fraga_2024,Cabrera-Vives_2024,Shah_2025,Gupta_2025,morenocartagena2025leveragingpretrainedvisualtransformers}.
Model accuracy was a key focus of the Kaggle competition and remains
a primary focus. However, it is not enough to have a high F1 score;
in order to be useful, a model needs to adapt to real LSST data, be
fast enough for large datasets, achieve strong performance for early
detection, provide useful embeddings on out-of-domain classes, and
give good uncertainty estimates.

Machine learning use for transient classification has a long history
\citep{Bailey_2007,Bloom_2012_aut_disc,Long_2012,Lochner_2016,Charnock_2017,Naul_2018,Carrasco-Davis_2019,Gómez_2020,Jamal_2020},
including recent efforts to classify SN types and variables \citep{Muthukrishna_2019,Villar_2019,Möller_2019_SuperNNova,Boone_2021,Qu_2021,Gagliano_2023,Pimentel_2023,Donoso-Oliva_2003_astromerv1,Rehemtulla_2024,Rizhko_2024}.
Very recent efforts include \citet{Li_2025} who explore using models
such as Moirai \citep{Woo_2024} and Chronos \citep{Ansari_2024},
finding good performance from time series models, despite those models
being trained on data from much different domains. \citet{Tan_2025}
and \citet{Donoso-Oliva_2025} apply transformer-style models to the
MACHO LC classification task. \citet{Vidal_2025} applies a transformer-based
model to synthetic MOSFiT light curves, estimating simulator parameters.

Machine learning systems often incur problems with domain adaptation
(please see \citealp{Koh_2020} for an excellent overview) and poor
uncertainty estimates, which can mean that their accuracy comes at
the expense of these secondary objectives. These problems are well-known,
but there is no scientific or industrial consensus on how to address
them; practitioners must examine a variety of methods, each with their
own drawbacks. Significant discrepancies between \elasticc and observational
data are expected. The effective domain for the problem may gradually
shift over time as well, as telescope conditions and scientific understandings
of the distribution/classification of various objects change, and
may benefit from techniques such as active learning \citep{Richards_2012}
or online learning (see \citealt{Mohri_2012} for ML theory). Machine
learning systems also exhibit issues with generalization (a subtly
different problem than domain adaptation, defined as lower accuracy
with a new sample from the same domain, commonly seen as divergence
in training/validation performance), and calibration. But at least
with \elasticc, these are more easily addressed with standard techniques,
as we demonstrate later.

We develop and evaluate several techniques to improve accuracy, early
detection, calibration, and label efficiency. We developed a light
curve encoding technique that significantly increases classifier accuracy
over previous state of the art. We leverage our decoder architecture
to train towards good early detection performance. We effectively
calibrate our model, and discuss the benefits of calibration while
cautioning that it does not resolve model bias. We do this while effectively
leveraging consumer-grade hardware for fast and cost-effective training
and inference. We also achieve strong performance when removing labels
from 90\% of the \elasticc dataset, demonstrating label-efficient
training. This should make our model useful for application to other,
smaller datasets, as well as facilitate use of techniques such as
active learning for ameliorating domain adaptation issues as these
models are applied to LSST observational data.

We experiment with pretraining via a generative objective, predicting
the next flux value at a given time and wavelength (and flux\_err).
This is similar to the objective used by NLP models such as langauge
models. The usefulness, bias, quality, and ethical implications of
LLMs are contested \citep{Bender_Gebru_Stochastic_Parrots,Lee_2023,perrigo2023openai,Raji_2022},
but they nevertheless exhibit a surprising level of generalization
given their pretraining objective (predicting the next word from previous
ones), and this pretraining is absolutely essential to model quality.
Moirai \citep{Woo_2024} has demonstrated that unsupervised learning
can be effectively leveraged for time-domain data, to produce better
forecasting estimates than several alternatives. In astronomical contexts,
\citet{Zhang_2024,Rizhko_2024} both use the CLIP constrastive objective
\citep{Radford_2021_CLIP} to align spectra and light curves, with
Zhang focusing on transients and Rizhko and Bloom focusing on variables.
These objectives are unsupervised and quite valuable, but geared towards
high-accuracy simulation data, whereas our pretraining attempts to
learn from the light curves directly, and we intend to pretrain on
{[}unlabeled{]} LSST observational data once that is available. \citet{Gupta_2025,Gupta_2025_SBPT}
explore transfer learning from a simulated ZTF dataset to the simulated
\elasticc dataset. While this exercises reuse of deep neural network
embeddings, it does not use an unsupervised objective. 

The LSST will produce vast quantities of unlabeled data (and previous
efforts such as the ZTF already have), and making use of it is desirable.
In addition to some accuracy boost, unsupervised pretraining can improve
generalization and domain adaptation \citep{why_pretrain_dl_2010_erhan_bengio,Goyal_2021}.
While all observational data is biased by which objects are bright
enough to be detectable, unsupervised data should have less selection
bias than labeled datasets. Unsupervised pretraining should also make
embeddings more generally useful, since the objective incentivizes
the last layer embedding to maintain information that may not matter
for classification. Our setup also allows for generative modeling.
While in astronomical contexts we have physical simulators which generate
quite meaningful light curves, our model may be useful for reproducing
observational phenomena that are not yet well captured by the simulations,
and providing an alternative preprocessing method for interpolation
to a new time / channel grid.

In this work, we introduce \nicknameWithEmoji, a classification
model that considerably improves on previous state of the art. In
many ways, it builds on the excellent work of ATAT \citep{Cabrera-Vives_2024}.
Our code and models are currently released at \url{https://atcat.click}.
In Section 2, we describe our methods. In Section 3, we present the
results of our model, including ablations that help explain which
of its improvements were most significant. In Section 4, we re-articulate
the scientific value of our model, and discuss many future directions
which could be taken with this work. For readers interested in the
encoding method which provides the largest \elasticc F1 scores improvement,
key results are presented in Section \ref{subsec:Methods-input-encoding}
and \ref{subsec:Input-encoder-ablation}.

\section{Methods}

\begin{figure}
\begin{centering}
\includegraphics[width=7cm]{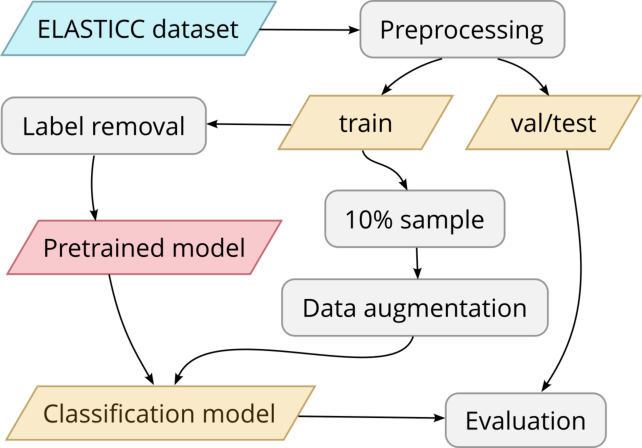}
\par\end{centering}
\caption{\textbf{}High-level schematic of dataflow for pretraining and fine-tuning
(training of the classification model). In implementation, all grey
nodes except preprocessing are run at runtime, with instrumentation
to double-check their correctness. See text.}\label{fig:Pipeline-schematic}
\end{figure}

In this section, we present our approaches for modeling photometric
light curves, primarily for the end goal of classification. A photometric
light curve consists of a sequence of observations, each consisting
of a time, channel (band pass color filter), flux error (provided
by instruments and/or upstream models), and a “calibrated flux”
value. This flux value has mean field subtraction applied, which for
our purposes means that it can be negative.

For classification purposes, our model takes this light curve as input
and produces a weighted distribution of class labels. For generative
and data augmentation purposes, we generally think of time, channel
(band pass color filter), and flux error values as being arbitrary
and given to us, but the flux reflecting the true nature of the object.
We consider its conditional distribution $P_{t,c}$ as a Gaussian,
with standard deviation given by flux error,
\begin{eqnarray}
\text{flux\_err}_{t,c} & \sim & P_{\text{err}}\ \text{(arbitrary distribution across the dataset)}\nonumber \\
\text{flux}_{t,c} & \sim & \mathcal{N}\left(\text{true\_flux}_{t,c},\text{flux\_err}_{t,c}^{2}\right)=:P_{t,c}\label{eq:conceptual-flux-err-dist}
\end{eqnarray}
For all models (except ablation studies), we replace the integer channel
index (representing 6 color filters from ultraviolet to infrared,
called u, g, r, i, z, Y) with the central wavelength of each color
filter / band pass. We will refer to this as “channel wavelength”.

An overview system diagram for the unsupervised pretraining and fine-tuning
setup is in \Figref{Pipeline-schematic}. The full-label training
dataflow is similar: simply replace the “10\% sample” node with
a pass-through and remove the “label removal” and “pretrained
model” nodes.

\subsection{The \elasticc dataset}

We train and evaluate our model on the \elasticc dataset, a realistic
simulation of LSST time-series data. Example light curves are shown
in \Figref{Local-attention} (more examples in the Appendix).

\subsubsection{ATAT splits and labels}\label{subsec:Methods-ATAT-splits}

We use the \elasticc v1 dataset \citep{elasticc_dataset_abstract_only}.
Our preprocessing is similar to ATAT \citep{Cabrera-Vives_2024} and
involves:
\begin{itemize}
\item The same 20-way classification scheme as ATAT, by combining some of
the original 32 classes. Please refer to \citep{Cabrera-Vives_2024}
Figure 2 for a class distribution frequency and motivation.
\item For each object (grouped by SNID), we find $t_{\text{alert}}$, when
the alert flag is present, and include observations (light curve points)
up to 30 days before this alert flag. Times are reset so that $t=0$
is the time of the first point.
\item The exact same training / validation / test splits as ATAT, as they
have shared the exact IDs used in each split. The data is first split
into a test distribution with 20,000 examples (1,000 for each class),
and the remainder as 5-way cross validation (so each split has 80\%
train and 20\% val).
\item For metadata, we train a \texttt{QuantileTransformer} with \texttt{output\_distribution=\textquotedbl normal\textquotedbl}
using scikit-learn, similar to ATAT. We save nan/null and $\pm\infty$
bits, and add separate embedding vectors when those bits are active.
\end{itemize}
Our preprocessing is very similar to ATAT to facilitate 1-1 comparisons,
although we use Polars for better CPU efficiency.  The only difference
in semantics is that for metadata, ATAT replaces nan/null/$\pm\infty$
values with -9999, whereas we preserve the semantic distinction between
these values, which may improve robustness when data has both missing
and extreme values.

\subsubsection{ Data augmentation}\label{subsec:Data-aug-methods}

To reduce overfitting to the training data, we implemented several
data augmentation processors. These routines were stacked upon one
another, so it was possible for multiple (or all) augmentation functions
to be applied to a single light curve. The routines include,
\begin{itemize}
\item Subsampling: selects a random subset of points, retaining at least
10 points. The exact number of points retained was chosen uniformly
between 10 and the number of points minus 1. It was randomly applied
at a rate of 25\%, although we passed through LCs with 10 or fewer
points unmodified, so its effective rate is a bit lower.
\item Flux scaling: scales \texttt{flux} and \texttt{flux\_err} by a uniform
random value in $\left[\frac{1}{1.1},1.1\right]\approx\left[0.909,1.1\right]$.
Each point in a light curve is scaled by the same value. It was randomly
applied at a rate of 20\%.
\item Time scaling: shifts the time of all points by a uniform random value
in $\left[0,10\right]$. Time shifting should help ensure the model
is robust if an alert is flagged a bit earlier or later. It was randomly
applied at a rate of 20\%.
\item Redshifting: modifies channel wavelengths. We sampled $z_{\text{additional}}\in\mathcal{N}\left(0,\sigma^{2}\right)$
with $\sigma=0.1$, and then clipped this to a min value of -0.1 (slightly
blue-shifted). Then we multiplied each channel wavelength by $\left(1+z_{\text{additional}}\right)$.
This augmentation was randomly applied at a rate of 20\%.
\item Random noise: samples an additional flux error term $e_{i,j}$ for
LC point $j$ of example $i$, and then samples the actual error $\xi_{i,j}\sim\mathcal{N}\left(0,e_{i,j}^{2}\right)$.
$e_{i,j}$ is added to \texttt{flux\_err}, and $\xi_{i,j}$ is added
to \texttt{flux}. $e_{i,j}$'s come from a scaled uniform distribution
$\text{Unif}\left[0.002\sigma,0.2\sigma\right]$, where $\sigma=\mathtt{std}\left(\text{flux}_{i}\right)$.
If we only have one flux value, we replace the standard deviation
term by 1. \emph{This augmentation is not applied to our main models},
but for all of our pretraining setups except the baseline, it was
randomly applied at a rate of 15\%.
\end{itemize}
Ablation studies can be found in Section \ref{subsec:Data-aug-results}.

\subsection{Model architecture}

\begin{figure}
\begin{centering}
\includegraphics[width=6cm]{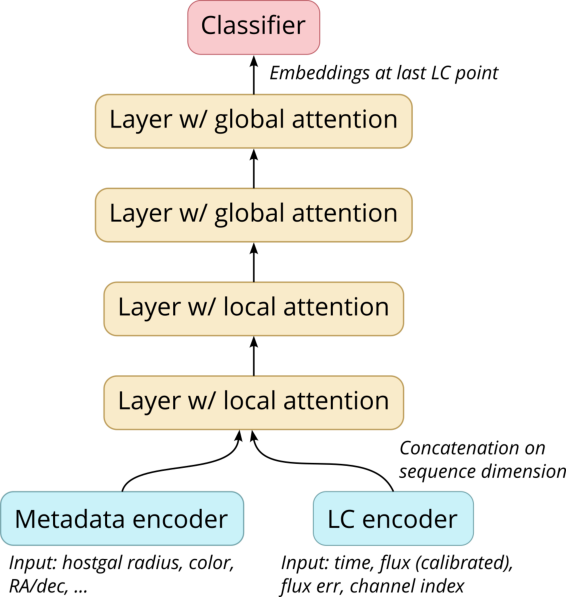}
\par\end{centering}
\caption{\textbf{Our model architecture}. Our model is a transformer with
4 layers. The first two layers use local attention, the first with
a 1-day threshold, and the second with a 10-day threshold. Metadata
is encoded as the first token in the sequence, allowing LC points
in the global attention layers to attend to it.}\label{fig:Architecture}
\end{figure}
Our model consists of encoders for metadata and light curves, a transformer
with 4 layers, and an output classifier, as shown in \Figref{Architecture}.
Our base architecture is a transformer, similar to \citet{Vaswani_2017}.
As ATAT \citep{Cabrera-Vives_2024} has demonstrated, transformers
are quite capable models, outperforming models based on feature extraction
and random forests (RF). Many other architectures have been attempted
(see Introduction), including convolutional networks and RNNs. The
transformer is popular in part because of the attention mechanism's
ability to effectively share information at various time scales, and
in part because it is relatively GPU-friendly.

We follow ATAT's settings for the main model dimension, and for the
rest of this section we denote
\[
d=\text{model dimension (default 384)}
\]
We used 4 attention heads, and a relatively small 64-dim attention
(so 16-dim key / query / value vectors per attention head). We omit
bias terms in key / query / value attention projections. Our feed-forward
units in each transformer layer had twice the embedding dimension
(768). Model sizes are the result of light tuning.

\subsubsection{ Light curve encoding}\label{subsec:Methods-input-encoding}

One major challenge of dealing with irregularly-sampled data like
\elasticc is encoding the values in an efficient manner. Convolutions
at various time-scales are a popular choice. Some of these like the
fourier transform involve basis functions over all time values, and
others such as wavelets or convolutional neural network kernels, use
windowed basis functions.

Following ATAT, we attempt to embed each LC point as a single sequence
element in the transformer. Let $i$ index into examples in our dataset
and $j$ index into the LC points of the $i$th example (there are
differing numbers of points per example), then
\[
x_{i,j}\in\mathbb{R}^{4}\sim\text{time},\text{flux},\text{flux\_err},\text{channel wavelength}
\]
Each of these 4-dimensional vectors will be embedded into a $d$-dimensional
vector.

Information is preserved if we can embed unique $x_{i,j}$ samples
to unique embeddings. Our approach has advantages compared to convolutional
or interpolation approaches, which might either choose a grid that
is too fine or too coarse. Grids that are too coarse can lose information.
Grids that are too fine add extra sequence elements, which in our
model would end up causing an $\mathcal{O}\left(n^{2}\right)$ cost
in the attention mechanism. Furthermore, it is common for examples
to be lacking measurements in one channel wavelength (this occurs
for \textasciitilde 18\% of \elasticc examples), and much less likely
but still possible that they only have measurements in a single channel
wavelength. Our approach appears better here, since it does not introduce
noisy estimations for these missing values.

A key concern is whether the network is actually able to extract usable
information from the embeddings, i.e. learn to compute reasonable
functions from it. We found that that the large set of dynamic (LC-based)
features extracted by ATAT no longer improved ATCAT performance, validating
the legitimacy of our approach, and suggesting that complex feature
extraction is not necessary for \elasticc classification.

Our input encoding routine consists of the following steps,
\begin{enumerate}
\item We embed each time value $t\in\mathbb{R}$ into a $d$-dimensional
vector
\[
\text{concat}\left(\left[\sin\left(\alpha_{k}t\right)\right]_{k},\left[\cos\left(\alpha_{k}t\right)\right]_{k}\right)
\]
Letting $m=\nicefrac{d}{2}$ and $k\in0..m-1$, we define $\alpha_{k}=10^{-T_{\text{max}}\cdot\beta_{k}}$,
where $\beta_{k}$'s are essentially the concatenation of two $m/2$-dimensional
vectors \texttt{linspace(-0.1, 0)} and \texttt{linspace(0, 1)}, except
without repeating the point at 0. In this manner, our $\alpha_{k}$
term is very similar to the term from \citet{Vaswani_2017}, but instead
of taking $\beta_{k}$'s as $\left[0,\nicefrac{1}{m},...,\nicefrac{m-1}{m}\right]$,
we choose time-scales from $-0.1$, capturing higher-frequency signals.
Indeed, $\Delta t$ between a pair of points in \elasticc can be
much smaller than 1, and so it makes sense to have some frequencies
(in our case, half of them) $\alpha_{k}>1$ where these $\sin$ and
$\cos$ values will be separated. By contrast, \citet{Vaswani_2017}
embeds positional indices in place of our $t$ values, which are integers
(their “$\Delta t$” is 1). We use $T_{\text{max}}=1500$ following
ATAT.
\item We scale the time values by $1/10$ to keep their $L_{2}$ norm around
1, and then pass them through a linear transformation.
\item We embed \texttt{flux} and \texttt{flux\_err} with a float-value embedder.
Each value $x\in\mathbb{R}$ is mapped to a small 4-dimensional vector
$\left[\tanh\left(s_{k}x\right)\right]_{k}$ where $s_{k}\in\left[1,10,1000,100000\right]$,
and then this is mapped through a \texttt{nn.Linear} layer to the
embedding dimension (384 by default). The \texttt{nn.Linear} (affine)
layer does have a bias term, but it is zero-initialized. We chose
these functions so that the network could effectively have terms that
vary linearly within a certain dynamic range. Our multi-scale tanh
approach allows us to preserve dynamic range while avoiding outlier
saturation of activations, which residual networks have difficulty
recovering from.
\item We rotate all of the time embeddings using the rotary position encoder
\citet{Su_2021}. We choose scales $\alpha_{i}$ as
\[
\log\alpha_{i}=4\sin\left(\frac{2\pi i}{d}-0.5\right)^{10}-4\qquad i\in0..\mathtt{model\_dim}-1
\]
We chose this analytic function visually, such that we could experiment
with rotations by both wavelength and time (shifting the 0.5 offset),
but our final models do not use this. We do not believe it is better
than the default rotary position encoder. We normalize the input “index”
(here, channel wavelength) to effectively rotate each pair of values
by $\alpha_{i}\frac{\lambda}{Z}$ where $\frac{\lambda}{Z}\in\left[0,1\right]$
except when redshifting beyond LSST wavelengths. Empirically, we found
that rotating only flux or flux\_err embeddings was worse, and rotating
all channels is about the same. We hypothesize that key-query comparisons
in the attention are gathering information on different time-scales,
and these rotations allow them to gather such information with a more
or less strong preference for matching the channel wavelength.
\end{enumerate}
Ablation studies can be found in Section \ref{subsec:Input-encoder-ablation},
future work in \ref{subsec:FW-modeling}.

\subsubsection{Metadata encoding}

We encode metadata by feeding the quantile scores for 85 features
into a basic embedder, consisting of a base $d\rightarrow d$ affine
transformation, and a residual nonlinearity consisting of a $d\rightarrow d$
affine transformation, leaky ReLU, and another $d\rightarrow d$ affine
transformation. Both affine transformations connected to the input
are given a slight L1 penalty, to encourage the model to ignore features
which do not provide a strong signal.

\subsubsection{Packing input sequences}\label{subsec:Packing-input-sequences}

For training classification models, we packed the metadata as a single
input token at the beginning, and then added embeddings for light
curve points. The index of the last light curve point is used to index
into the last layer for classification, using \texttt{torch.gather},
because each element in a batch of light curves may have a different
sequence length.

Encoding values for our pretrained model was more interesting. After
each light curve point, we included the time, channel wavelength,
and flux error for the next point. One batch element might look like:

\texttt{{[}META{]} {[}LC0{]} {[}PRED1{]} {[}LC1{]} {[}PRED2{]} ...}

Then, for training the generative model, we again used a gather operation
to retrieve all of the \texttt{{[}PREDICT{]}} sequence elements. In
terms of implementation, generating these indices can be tricky; we
found it effective to use sequential CPU code (per batch element),
since the total amount of computation is small compared to the model.

\subsubsection{Transformer with local attention}\label{subsec:Methods-Local-attn}

\begin{figure}
\begin{centering}
\includegraphics[width=6.7cm]{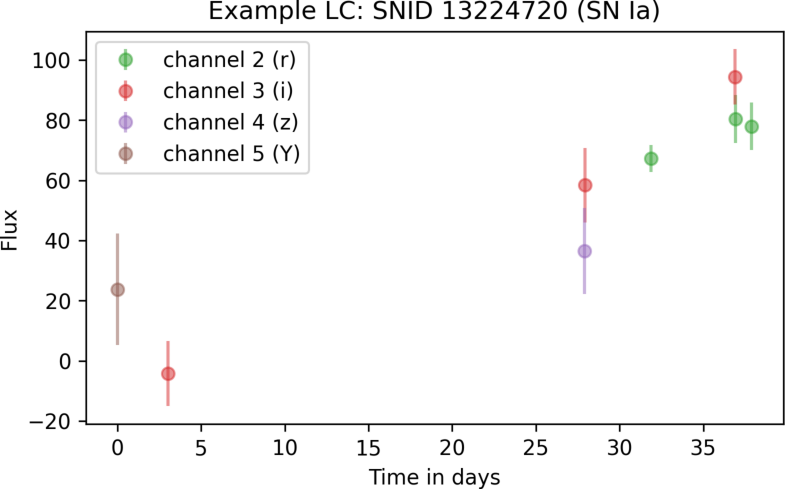}
\par\end{centering}
\vspace{0.3cm}
\begin{centering}
\includegraphics[width=7cm]{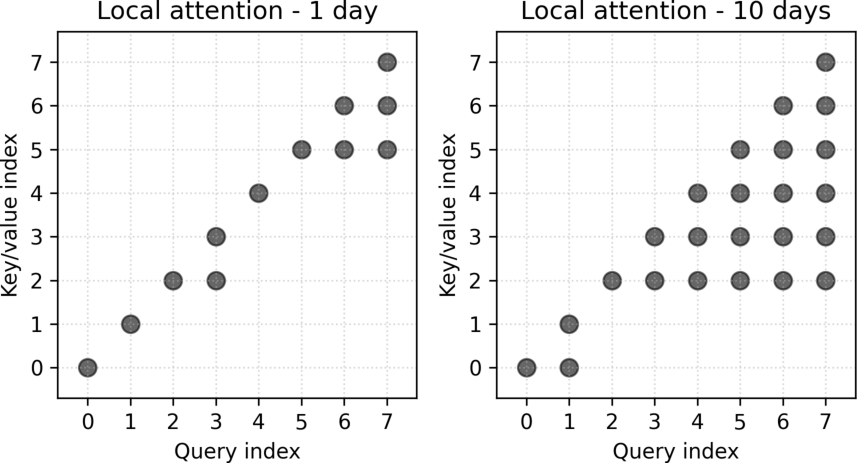}
\par\end{centering}
\caption{\textbf{Local attention connections}. We visualize the connections
in our local attention mechanism for a specific \elasticc example
(this one is a Type-Ia supernova). The first transformer layer, featuring
a local attention mechanism with a threshold of 1 day, allows the
two points in channels 3 (\texttt{i}) and 4 (\texttt{z}) near $t=27$
to attend to each other, and likewise the cluster of 3 points at the
end (near $t=37$), but only causally (bottom left figure). For the
first pair, this is represented by the point at query index 3 being
allowed to attend to key/value index 2 (presence of a black dot).
For the second layer, with a threshold of 10 days (bottom right figure),
the first two points can attend to each other, and all of the other
points as well, but also only causally.}\label{fig:Local-attention}
\end{figure}

Our main transformer stack, shown in \Figref{Architecture}, is a
4-layer hybrid transformer. The first two layers have a local masking
scheme in their attention mechanism. This means that each LC point
is only allowed to attend to points which are before it and within
a time range. For example, if our threshold is 10 (days), then a point
with $t=31$ can attend to a point with $t=25$ but not a point with
$t=20$. An example of this mechanism is in \Figref{Local-attention}.
For local transformer layers, sequence elements containing metadata
could not attend to sequence elements for any light curve points.
Our model is fully causal, also known as a decoder architecture. Causal
masking means that any light curve point can only attend to those
before it. We do not believe causal masking provided any performance
boost over full attention, but did make it easier to train early detection
and generative models.

\subsubsection{Classification}

We use a very simple classification layer, consisting of a layer norm,
and affine projection to the number of output classes. For most of
our models, we also added an extra output class, which we intended
to correspond to “unsure”. Scores were calculated as a softmax
(or often log-softmax) including this class. However, we found that
even with Soft F1 losses (see next section), the models generally
would only learn to put very minimal weight on this class, learning
to output a smoothed distribution over other classes. We did use this
extra output class for calibration (Section \ref{subsec:Methods-calibration})
and novel object detection (Section \ref{subsec:Preliminary-work-Anomaly}).

\subsection{Evaluation and losses}

For most of this work, we are reporting macro F1 scores. Please see
\citet{Opitz_2019} for a precise technical definition. In general,
for all classes, we will compute
\[
P_{i}=\frac{\#\text{TP (true pos)}}{\#\text{TP}+\#\text{FP (false pos)}}\qquad R_{i}=\frac{\#\text{TP}}{\#\text{TP}+\#\text{FN (false neg)}}
\]
and then, the harmonic mean of $P_{i},R_{i}$ and then take a simple
average over classes. For our evaluation metrics, we use the implementation
in \texttt{torcheval.metrics}.

\subsubsection{Class-balanced metrics for the val dataset}

We implemented a version of the F1 metric that is less sensitive to
class imbalances early on, and argue that it contributed greatly to
the integrity of our experiments. The val and test splits differ greatly
in their distributions of exemplars, and therefore using a metric
which in expectation is the same for both means that we didn't need
to look at any test metrics until this writeup.

Macro F1 alone does not fix class imbalance. Suppose we have $n$
examples of $P_{i}$ and $n$ examples of another class $P_{i'}$,
and the classifier confuses these classes at some rate. If we add
10x more elements of our “other” class $P_{i'}$, even if they
are from the same distribution, the FP count of $P_{i}$ will be greatly
affected. On \elasticc with ATAT's splits, this effect is very strong.
We provide a very concise explanation of our class-balanced F1 score,
loosely following the notation of \citet{Opitz_2019}. Formally, let
$X$ be the space of input features, $Y=1..n$ be our label set, $f:X\rightarrow Y$
be our classifier, $S$ be our dataset consisting of $\left(x,y\right)$
pairs, and $\left\{ m_{ij}\right\} _{i\in1..n,j\in1..n}$ be our weighted
confusion matrix,
\[
m_{ij}=\sum_{x,y\in S}w_{x,y}1_{f\left(x\right)=i\text{ and }y=j}
\]
where $w_{x,y}\propto\frac{1}{\left|\left\{ x',y'\in S:y'=y\right\} \right|}$
weights by inverse class frequency, and $1_{\text{cond}}$ is an indicator
function. Then define
\[
P_{i}=\frac{m_{ii}}{\sum_{y=1}^{n}m_{iy}}\qquad R_{i}=\frac{m_{ii}}{\sum_{y=1}^{n+1}m_{yi}}
\]
and proceed as normal.

\subsubsection{Soft/trainable F1 scores}

We also implemented a version of soft F1 scores for training. These
can be defined by considering our model's outputs as a vector of scores
(formally, $f:X\rightarrow\left[0,1\right]^{Y}$), and writing
\[
m_{ij}=\sum_{x,y\in S}w_{x,y}f\left(x\right)_{i}1\left\{ y=j\right\} 
\]
(In our training setups, we don't need to worry about the weight term,
because we use a balanced batch sampler.)

\subsection{Training routine}\label{subsec:Methods-training-routine}

We used Pytorch default initializers (Kaiming for linear/affine transformations),
except for float value embedders as mentioned previously. Our models
which had full access to training data did not have dropout, but for
ensembled models for smaller labeled sets (our unsupervised pretraining
experiments), we used 15\% dropout for transformer layers, and 4\%
for input and classifiers. We did not see much difference whether
dropout was placed before or after each attention/feed-forward unit,
and have them after by default.

We used a training regime similar to ATAT, using an nadam optimizer
with 2e-4 learning rate, $\beta_{1}=0$ and $\beta_{2}=0.999$, which
is similar to RMSProp. We used a learning rate schedule with a linear
warm-up for the first 1000 steps, steady 2e-4 learning rate, then
exponential decay after 20,000 steps with a half-life of 6,000 steps.
For pretraining, we randomly shuffle training data; for fine-tuning
/ labeled training, we used ATAT's balanced batch sampling approach,
which yields batches that are balanced among labels. For setups with
less data, we enabled the nadam weight decay term with weight 1e-5
as well. During pretraining, we clip gradients to a L2 norm of 0.1,
but we do not have any gradient clipping for fine-tuning / labeled
training.

For the setup with the full training dataset, we trained for 40,000
steps, evaluating every 4,000 steps. For the main results, we used
our class-balanced F1 metrics on the validation set to select the
ideal checkpoint, but for several sub-experiments such as early detection,
calibration, and generation of confusion matrices, we picked step
36,000 for implementation convenience. Generally, the model performance
would change by at most 0.14\% after step 32,000. For our unsupervised
pretraining setups, we pretrained models for 20,000 steps, and then
fine-tuned for 20,000 steps.

\subsection{Unsupervised pretraining}\label{subsec:Methods-unsupervised-pretraining}

\begin{figure}
\begin{centering}
\includegraphics[width=8cm]{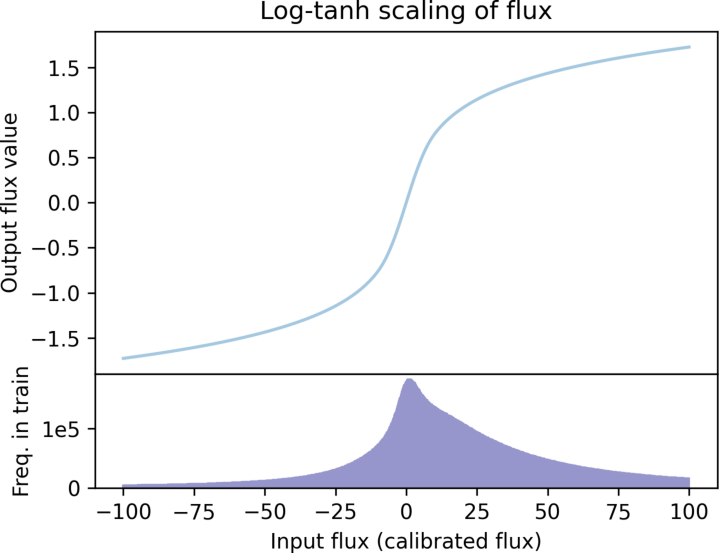}
\par\end{centering}
\caption{\textbf{Nonlinear scaling of flux values for generative modeling}.
We squash the flux values to a much smaller range, by “gluing”
together a tanh function (around 0) and log function, matching the
first derivative and intercept point. We also pre-scale flux by 1/10.
This keeps the response curve not too flat for the majority of values,
while scaling the max value from 2,568,897 to 6. The first part of
the figure is the response curve, the bottom is a histogram (aligned
in X-axis values) of all training flux values from all light curves
(not showing the long tail of extreme values). As elsewhere, we are
looking at calibrated flux values after mean field subtraction, which
can be negative.}\label{fig:Pretr-nonlinear-scaling}
\end{figure}
\begin{figure}
\begin{centering}
\includegraphics[width=8cm]{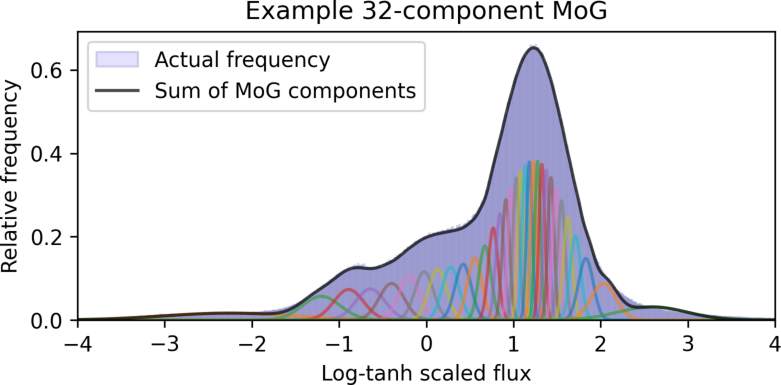}
\par\end{centering}
\caption{\textbf{Mixture-of-Gaussian components for generative modeling}.
We graph the components of our mixture of Gaussians model, showing
actual frequencies of scaled flux values, and the sum of Gaussian
components (black line, each component equally weighted). For our
actual models, we used 64 and 128-component models; here we show only
32 for ease of visualization. The black line matches our histogram
fairly well, as desired, with some deviation around \textasciitilde 2.4
for this 32-component model. Incorrect choices of sigmas (component
width) will result in the black line being jagged, or components being
too broad. Individual components are the many colored lines.}\label{fig:Pretr-mog-model}
\end{figure}

As we briefly explained in the introduction, our goal with unsupervised
pretraining is to capture information about the distributions of light
curves in the model parameters, so that it will generalize better
when trained on a limited amount of supervised data. In order to mimic
this setup for \elasticc, we removed labels from 90\% of the training
data. We found it much more convenient and equally performant to run
unsupervised pretraining on all training data, and then fine-tune
on the 10\% of training data with labels, although we also tried mixing
these objectives.

We experimented with two objectives:
\begin{itemize}
\item Predicting values: This is the dominant approach for LLMs. Unlike
LLMs, we chose to model the conditional distribution $p\left(\text{flux}|\text{time},\text{flux\_err},\text{channel wavelength}\right)$,
but not model the distribution of the subsequent time, channel wavelength,
or flux\_err values.
\item Patch prediction from full transformer attention: this approach was
used by Moirai \citep{Woo_2024}. We masked out contiguous ranges
of points between 5\% and 50\% of the total range of time values,
with hand-tuned logic to select ranges of data that appeared informative
(details omitted, since we did not proceed with this approach).
\end{itemize}
Preliminary experiments showed that the two approaches had similar
value for the purpose of fine-tuning classification models. The first
approach is simpler and more data efficient, so we ended up using
it. Please see Section \ref{subsec:Packing-input-sequences} for how
we formatted these inputs to our model; for the \texttt{{[}PREDICT{]}}
elements, we used the same embedding technique as Section \ref{subsec:Methods-input-encoding},
omitting the flux component and adding a free parameter (so that the
model could learn to distinguish token types).

\subsubsection{Nonlinearity}\label{subsec:Methods-Nonlinearity}

For unsupervised pretraining, we deal with extreme values in the \elasticc
dataset slightly differently than in feature encoding, since our approach
in feature encoding (of combining multiple scalings) cannot be used
here. We ended up squashing the target flux values with a nonlinearity
$f$. This function is shown in \Figref{Pretr-nonlinear-scaling},
and is similar to the nflows codebase \citep{nflows}. We projected
flux to $f\left(\text{flux}\right)$, and
\begin{equation}
\text{scaled flux\_err}=\frac{f\left(\text{flux}+\text{flux\_err}\right)-f\left(\text{flux}-\text{flux\_err}\right)}{2}\label{eq:flux_err_scaling}
\end{equation}
Recall that we are thinking of modeling the distribution of true flux
with observational error, $P_{t,c}$ as in (\ref{eq:conceptual-flux-err-dist}).
We don't have access to $\text{true\_flux}_{t,c}$ and want to work
in the projected space (with less extreme values), so we instead conceptualize
\[
P'_{t,c}=\mathcal{N}\left(f\left(\text{scaled flux}_{t,c}\right),\sigma^{2}=f\left(\text{scaled flux\_err}_{t,c}\right)^{2}\right)
\]
This makes little sense statistically, and we think that experimenting
with Moirai's approach of using heavier-tailed distributions (e.g.
t-distributions) would be valuable future work (see \citealp{Woo_2024}).
For the goal of getting useful hidden embeddings for downstream use,
how principled we are may not be of utmost importance. For generative
modeling, it is more important; our model won't fail to run, but it
won't capture several phenomena that it should.

We expressed our predicted next flux distribution $Q_{t,c}$ as mixture
weights for a Mixture-of-Gaussians model, where the means and sigmas
are fixed. These fixed values come from the training data with a very
simple algorithm; we chose our $n$ means and sigmas as
\begin{align*}
\vec{\text{means}}\left(n\right) & =\mathtt{quantile}\left(\vec{\text{flux}},\mathtt{linspace}\left(\nicefrac{1}{2n},1-\nicefrac{1}{2n},n\right)\right)\\
\vec{\text{sigmas}}\left(n\right) & =0.7\cdot\mathtt{diff}\left(\vec{\text{means}}\left(n+1\right)\right)
\end{align*}
where the monospace-font functions are from torch or numpy. Visually,
this seemed to match our distribution well; see \Figref{Pretr-mog-model}.
We also tried tuning the means and sigmas by maximizing the log-likelihood
of samples of flux values, but found that this only increased noise.

In order to generate a predicted next flux distribution in the original
space, we can project the means and standard deviations back, since
our nonlinearity $f$ is invertible. We use the same approach for
flux\_err as (\ref{eq:flux_err_scaling}), simply replacing $f$ with
$f^{-1}$. Our approach fails to represent the most extreme points.
However, it solved a practical problem we encountered trying to predict
unscaled flux values: the model would often learn to use the very
wide mixture components at the extremes rather than ones with much
closer means, because the Gaussian tails fall off too quickly. We
hypothesize that Moirai's approach would fix this issue.

\subsubsection{Predicting flux with observational noise}

Since the values we are trying to predict in \elasticc include observational
error, we have a few natural choices for how to evaluate our model.
Do we care about forward KL divergence,
\[
\text{KL}\left(P'_{t,c}||Q_{t,c}\right)
\]
or reverse KL, $\text{KL}\left(Q_{t,c}||P'_{t,c}\right)$? Our insight
should derive from the fact that for distributions $A$ and $B$,
if $B$ is broader then $A$, then
\[
\text{KL}\left(A||B\right)<\text{KL}\left(B||A\right)
\]
For reverse KL, when the model is quite confident (say, it just got
a low-noise LC point in the same band), then it won't get harshly
penalized by a point with high observational noise. However, when
observational noise is low, but there just isn't enough information,
then the model will get a huge penalty. In practice, these huge penalties
end up overtaking the loss, and we have to artificially soften $P'_{t,c}$
to learn anything at all.

But forward KL is not perfect either, especially in the case we just
mentioned when the model is confident and correct, but we have a lot
of observational noise. The same applies to logloss / $\delta_{\text{flux}}$.
Our solution is to instead think about modeling
\[
Q_{t,c,\text{err}}\sim\text{flux}_{t}|t,c,\text{flux\_err}_{t,c},\text{(previous points)}
\]
and the model can therefore learn to increase its variance if we have
a high $\text{flux\_err}_{t,c}$.

There is a third possibility, log-loss “$q_{t,c}\left(\text{flux}_{t,c}\right)$”,
denoting the density $q_{t,c}$ {[}which can also be conditioned on
error, for $q_{t,c,\text{err}}${]}. By taking limits or relaxing
formalities appropriately, this is equal to $\text{KL}\left(\delta_{\text{flux at }t,c}||Q_{t,c}\right)$
where $\delta_{\text{flux at }t,c}$ is the Dirac delta, similar to
forward KL. It appears to not assume $P'_{t,c}$ is normal, which
could be a considerable advantage, at the cost of potentially being
less efficient, because it will only learn the variance of $P'_{t,c}$
from seeing many samples over our dataset, instead of directly from
$\text{flux\_err}_{t,c}$. We leave experimentation as future work.

\subsubsection{Generative performance evaluation}

In Section \ref{subsec:Results-pretraining}, we compare generative
performance with the Gaussian Process used by \citet{Boone_2019},
projecting our distributions of predicted flux back to the original
unscaled space, as described in Section \ref{subsec:Methods-Nonlinearity}.
We studied forward KL divergence on 1000 test set examples, by first
filtering out sequences with less than 6 points, and then randomly
selecting 50 from each class. This was a smaller study partly because
the GP was expensive to run in an iterative-decoding fashion, since
it needs to be re-trained for each new LC point. For both models,
we start measuring prediction performance on the 5th point; this resulted
in predictions on 61,099 LC points in total. Since the Gaussian process
was only predicting a single Gaussian, we used the simpler closed
form for its KL divergence. For our model, we had to sample the KL
divergence, and drew 4000 samples for each point. To ensure that we
had drawn enough samples to accurately compute the KL, we subsampled
this sample (using 5-way CV-like folds), and ensured that the standard
deviation of subsample KL scores was small (its median is 0.005).

\subsection{Early detection}\label{subsec:Methods-Early-detection}

In order to improve early detection performance, we simply added a
logloss which is evaluated at many LC points instead of just the final
one (we skipped the first point on the assumption it may be too noisy).
We averaged this loss across sequence length per example first, so
that long examples would not get disproportionately more loss than
short ones. We noticed that final prediction accuracy decreased with
only this loss, and so we compensated by adding back our old last-point
loss (in the form of logloss, not Soft F1) with 1/3 of the early detection
loss weight. This approach makes good use of our causal decoder architecture,
effectively training towards early detection performance at all time
values simultaneously.

\subsection{Calibrated models}\label{subsec:Methods-calibration}

We evaluated several methods to calibrate our model, including
\begin{enumerate}
\item Simple label smoothing, following Section 7 of \citet{Szegedy_2015}
\item Focal loss \citep{Lin_2017}, including a variant where we removed
the loss re-weighting from the gradient calculation
\item Training the unsure logit towards inverse train-time accuracy on each
example
\item Training a bootstrap ensemble, and then training the unsure logit
towards the bootstrap ensemble's accuracy on each example
\end{enumerate}
Label smoothing adds a $\nicefrac{\epsilon}{n}$ to the true probabilities
for each example (subtracting from the label class' probability);
we choose $\epsilon=0.1$ following \citet{Szegedy_2015}.

Focal loss \citep{Lin_2017} re-weights examples by their difficulty.
For a single example, suppose $q_{\text{true}}$ is the output probability
of the true class. 
\[
w\left(q_{\text{true}}\right)=\left(1-q_{\text{true}}\right)^{\gamma}\qquad\text{FL}\left(q_{\text{true}}\right)=-w\left(q_{\text{true}}\right)\log\left(q_{\text{true}}\right)
\]
We tried $\gamma\in\left\{ 0.5,1,2\right\} $. We tried detaching
calculations in $w$ from automatic gradient calculation, and found
that this consistently provided a slight improvement. For our setup,
$\gamma=0.5$ seemed sufficient to get good calibration results, and
had the best F-1 scores.

We also tried a runtime technique where we try to have the “unsure”
logit (see Section \ref{subsec:Methods-Local-attn}) probabilities
to correspond to accuracy of each batch at training time (to compute
accuracy, we compute the mean over batch elements where the logit
with the highest output probability is correct, detaching tensors
from automatic gradient calculation). Let $b$ index over batch elements,
and $q_{b,\text{true}}$ be the output probability of the true class
for batch element $b$. We first compute a smoothed “confident
and correct” score, and from this create an “unsure” score
for each batch element,
\begin{align*}
\mathtt{cc}_{b} & =\text{avg}\left(q_{b,\text{true}},\text{min}\left(0,2\left(q_{b,\text{true}}-\nicefrac{1}{2}\right)\right)\right)\\
\mathtt{unsure}'_{b} & =\frac{0.1}{N_{\text{classes}}}+\left(1-\mathtt{cc}_{b}\right)\\
\mathtt{unsure}_{b} & =\frac{1-\mathtt{accuracy}}{\text{mean}_{b}\left(\mathtt{unsure}'\right)}\mathtt{unsure}'_{b}
\end{align*}
and then smooth the actual label probabilities, so the true label
gets $1-\mathtt{unsure}_{b}$ weight and the unsure label gets $\mathtt{unsure}_{b}$
weight; we train towards this with KL divergence. Effectively, we
train the model so that it will output “unsure” at a rate corresponding
to its accuracy, but allowing it to be quite confident on some examples
and less confident on others.

Finally, we tried a bootstrapping method, using a first round of classifiers
to estimate how difficult each example is. We first train 4 boostrap
sub-models, each of which gets 1/2 of the data. We minimize the maximum
overlap between these models, a fun mini mathematical puzzle, where
one divides the training into 6 partitions and assigns 3 of these
to each model. We intentionally under-train this for 12k steps and
regularize at fairly high 20\% dropout, 1e-4 weight decay, and then
ensemble these models together (adding probability scores) and run
it on the training to generate per-class probabilities for each class.
We do keep all sub-models in the ensemble, so half of them will have
seen a particular example as training input; given the early stopping
and high regularization, we presume that it hasn't overfit terribly.
We take the probability of the true class as our $p_{\text{true}}$
and one minus this as our $p_{\text{unsure}}$, and add a loss to
train the classifier's logits towards this using KL divergence. We
keep the base soft F1 loss towards just the true label to lose too
much overall quality.

\section{Results}

\subsection{Main results}\label{subsec:Main-results}

\begin{table*}
\begin{centering}
\begin{tabular}{|c|c|c|c|c|c|c|}
\hline 
Model & Architecture & \# params & Dataset (\# train/val/test) & \# classes & LC-only & LC+meta\tabularnewline
\hline 
\hline 
ATAT & transformer & 1.9M & \elasticc v1 (\res{our train size}/\res{our val size}/\res{our test size}) & 20 & \res{atat lc_only f1}\% F1 & \res{atat lc_meta f1}\% F1\tabularnewline
\hline 
ORACLE & RNN &  & \elasticc v2 (449K/225K/23K) & 19 & - & 84\% F1\tabularnewline
\hline 
SWINv2-based & vision transformer & 29M & Exact same as ATAT & 20 & 65.5 $\pm$ 0.28\% F1 & -\tabularnewline
\hline 
Ours & transformer & \res{our lc_meta num params} & Exact same as ATAT & 20 & \textbf{\res{our lc_only f1}\% F1} & \textbf{\res{our lc_meta f1}\% F1}\tabularnewline
\hline 
\begin{cellvarwidth}[t]
\centering
Ours, unsup.

pretrain
\end{cellvarwidth} & \begin{cellvarwidth}[t]
\centering
ensemble of

transformers
\end{cellvarwidth} & \begin{cellvarwidth}[t]
\centering
\res{our lc_meta num params}

/ model
\end{cellvarwidth} & \begin{cellvarwidth}[t]
\centering
Unsupervised pretraining: 1.5M

Supervised: 145K/\res{our val size}/\res{our test size}
\end{cellvarwidth} & 20 & \res{pretr setup A lc_only f1}\% F1 & \res{pretr setup A lc_meta f1}\% F1\tabularnewline
\hline 
\end{tabular}
\par\end{centering}
\caption{\textbf{Results from several state-of-the-art models on \elasticc}.
All $\pm$ values are simple \texttt{np.std()} calculations on the
F1 metric per CV fold. We only ran our unsupervised pretraining experiment
on the first 3 CV folds; others use all 5 folds. See text for details.}\label{tab:Model-comparison}
\end{table*}

\begin{table*}
\begin{centering}
\begin{tabular}{l|ccccccccc}
\hline 
 & \multicolumn{3}{c|}{RF (MD + Features)} & \multicolumn{3}{c|}{ATAT (LC + MD + Features)} & \multicolumn{3}{c|}{ATCAT (LC+metadata)}\tabularnewline
\hline 
Class & Precision & Recall & F1 & Precision & Recall & F1 & Precision & Recall & F1\tabularnewline
\hline 
CART & 59.2 $\pm$ 0.4 & 56.2 $\pm$ 0.6 & 57.6 $\pm$ 0.5 & 75.3 $\pm$ 2.5 & 40.0 $\pm$ 4.6 & 52.0 $\pm$ 3.3 & \textbf{78.8 $\pm$ 1.7} & \textbf{73.2 $\pm$ 1.0} & \textbf{75.8 $\pm$ 0.5}\tabularnewline
Iax & 57.6 $\pm$ 0.5 & 55.9 $\pm$ 0.6 & 56.8 $\pm$ 0.5 & 59.8 $\pm$ 2.1 & 65.1 $\pm$ 5.4 & 62.2 $\pm$ 1.9 & \textbf{72.7 $\pm$ 1.0} & \textbf{80.9 $\pm$ 0.9} & \textbf{76.6 $\pm$ 0.9}\tabularnewline
91bg & 75.2 $\pm$ 0.4 & 90.2 $\pm$ 0.2 & 82.0 $\pm$ 0.2 & 88.8 $\pm$ 2.2 & 92.5 $\pm$ 1.9 & 90.5 $\pm$ 0.6 & \textbf{93.5 $\pm$ 0.6} & \textbf{96.2 $\pm$ 0.3} & \textbf{94.9 $\pm$ 0.3}\tabularnewline
Ia & 61.4 $\pm$ 0.4 & 76.7 $\pm$ 0.2 & 68.2 $\pm$ 0.3 & 76.3 $\pm$ 1.2 & 81.4 $\pm$ 1.7 & 78.8 $\pm$ 0.7 & \textbf{79.7 $\pm$ 0.5} & \textbf{86.9 $\pm$ 0.4} & \textbf{83.1 $\pm$ 0.3}\tabularnewline
Ib/c & 58.0 $\pm$ 0.3 & 39.6 $\pm$ 0.4 & 47.1 $\pm$ 0.2 & 50.0 $\pm$ 3.8 & \textbf{65.8 $\pm$ 3.4} & 56.6 $\pm$ 1.2 & \textbf{69.6 $\pm$ 0.8} & 63.0 $\pm$ 0.9 & \textbf{66.1 $\pm$ 0.2}\tabularnewline
II & 66.8 $\pm$ 0.6 & 42.7 $\pm$ 0.5 & 52.1 $\pm$ 0.5 & 63.9 $\pm$ 3.5 & 66.4 $\pm$ 2.8 & 65.0 $\pm$ 1.3 & \textbf{75.3 $\pm$ 0.3} & \textbf{74.6 $\pm$ 0.5} & \textbf{75.0 $\pm$ 0.3}\tabularnewline
SN-like/Other & 59.0 $\pm$ 0.5 & 54.1 $\pm$ 0.8 & 56.5 $\pm$ 0.6 & 64.3 $\pm$ 2.2 & 60.5 $\pm$ 2.9 & 62.3 $\pm$ 1.5 & \textbf{70.6 $\pm$ 1.1} & \textbf{69.6 $\pm$ 1.1} & \textbf{70.1 $\pm$ 1.0}\tabularnewline
SLSN & 90.3 $\pm$ 0.1 & 90.0 $\pm$ 0.1 & 90.2 $\pm$ 0.1 & 89.6 $\pm$ 0.9 & \textbf{95.4 $\pm$ 0.4} & 92.4 $\pm$ 0.4 & \textbf{96.1 $\pm$ 0.3} & 94.2 $\pm$ 0.4 & \textbf{95.1 $\pm$ 0.3}\tabularnewline
PISN & 85.6 $\pm$ 0.1 & 96.7 $\pm$ 0.1 & 90.8 $\pm$ 0.0 & 95.9 $\pm$ 0.4 & 96.7 $\pm$ 0.9 & 96.3 $\pm$ 0.4 & \textbf{97.7 $\pm$ 0.2} & \textbf{98.4 $\pm$ 0.3} & \textbf{98.0 $\pm$ 0.2}\tabularnewline
TDE & 83.2 $\pm$ 0.4 & 76.8 $\pm$ 0.3 & 79.9 $\pm$ 0.2 & 79.0 $\pm$ 4.9 & \textbf{92.5 $\pm$ 1.0} & 85.1 $\pm$ 2.6 & \textbf{93.0 $\pm$ 0.9} & 92.2 $\pm$ 0.4 & \textbf{92.6 $\pm$ 0.5}\tabularnewline
ILOT & 76.3 $\pm$ 0.3 & \textbf{93.6 $\pm$ 0.2} & 84.1 $\pm$ 0.2 & 92.1 $\pm$ 0.9 & 84.0 $\pm$ 3.1 & 87.8 $\pm$ 1.3 & \textbf{92.7 $\pm$ 0.9} & 91.1 $\pm$ 0.3 & \textbf{91.9 $\pm$ 0.4}\tabularnewline
KN & 86.8 $\pm$ 0.2 & 90.3 $\pm$ 0.1 & 88.5 $\pm$ 0.1 & \textbf{97.1 $\pm$ 0.4} & 77.1 $\pm$ 2.5 & 85.9 $\pm$ 1.4 & 95.2 $\pm$ 0.4 & \textbf{94.6 $\pm$ 0.6} & \textbf{94.9 $\pm$ 0.1}\tabularnewline
M-dwarf Flare & 95.0 $\pm$ 0.3 & 79.4 $\pm$ 0.3 & 86.5 $\pm$ 0.3 & \textbf{99.1 $\pm$ 0.3} & 70.4 $\pm$ 1.9 & 82.3 $\pm$ 1.3 & \textbf{99.0 $\pm$ 0.3} & \textbf{89.5 $\pm$ 0.6} & \textbf{94.0 $\pm$ 0.4}\tabularnewline
uLens & \textbf{96.9 $\pm$ 0.4} & 82.8 $\pm$ 0.2 & 89.3 $\pm$ 0.3 & 86.8 $\pm$ 1.7 & 95.6 $\pm$ 0.7 & 91.0 $\pm$ 0.7 & 93.4 $\pm$ 0.4 & \textbf{95.8 $\pm$ 0.7} & \textbf{94.6 $\pm$ 0.2}\tabularnewline
Dwarf Novae & 78.5 $\pm$ 0.2 & 82.9 $\pm$ 0.3 & 80.6 $\pm$ 0.2 & 86.2 $\pm$ 1.8 & 92.0 $\pm$ 0.9 & 89.0 $\pm$ 0.9 & \textbf{91.7 $\pm$ 0.3} & \textbf{96.7 $\pm$ 0.2} & \textbf{94.1 $\pm$ 0.1}\tabularnewline
AGN & 95.4 $\pm$ 0.4 & \textbf{99.9 $\pm$ 0.1} & 97.6 $\pm$ 0.2 & 99.7 $\pm$ 0.1 & \textbf{100.0 $\pm$ 0.0} & 99.8 $\pm$ 0.1 & \textbf{100.0 $\pm$ 0.1} & \textbf{100.0 $\pm$ 0.0} & \textbf{100.0 $\pm$ 0.0}\tabularnewline
Delta Scuti & 90.8 $\pm$ 0.3 & 98.9 $\pm$ 0.0 & 94.7 $\pm$ 0.2 & 98.7 $\pm$ 0.3 & \textbf{99.5 $\pm$ 0.1} & 99.1 $\pm$ 0.1 & \textbf{99.3 $\pm$ 0.2} & \textbf{99.6 $\pm$ 0.1} & \textbf{99.4 $\pm$ 0.2}\tabularnewline
RR Lyrae & 91.6 $\pm$ 0.4 & 98.9 $\pm$ 0.1 & 95.1 $\pm$ 0.2 & \textbf{99.5 $\pm$ 0.2} & 99.1 $\pm$ 0.2 & 99.3 $\pm$ 0.1 & \textbf{99.6 $\pm$ 0.2} & \textbf{99.5 $\pm$ 0.1} & \textbf{99.5 $\pm$ 0.1}\tabularnewline
Cepheid & 92.6 $\pm$ 0.5 & 98.9 $\pm$ 0.1 & 95.6 $\pm$ 0.3 & 99.2 $\pm$ 0.3 & \textbf{99.5 $\pm$ 0.1} & 99.3 $\pm$ 0.1 & \textbf{99.5 $\pm$ 0.1} & \textbf{99.6 $\pm$ 0.2} & \textbf{99.6 $\pm$ 0.1}\tabularnewline
EB & 93.5 $\pm$ 0.3 & 97.5 $\pm$ 0.1 & 95.5 $\pm$ 0.2 & 90.4 $\pm$ 1.7 & 99.6 $\pm$ 0.1 & 94.8 $\pm$ 0.9 & \textbf{98.4 $\pm$ 0.3} & \textbf{99.7 $\pm$ 0.1} & \textbf{99.0 $\pm$ 0.1}\tabularnewline
\hline 
\end{tabular}
\par\end{centering}
\caption{\textbf{Fine-class comparison}. ATCAT has strong F1 scores on all
fine classes, while occasionally making a different precision-recall
tradeoff than the RF model or ATAT. $\pm$ values are from \texttt{np.std}
over 5 CV folds.}\label{tab:Fine-class-results}
\end{table*}

In Table \ref{subsec:Main-results} we show the main results for state
of the art \elasticc classification models. The third model, a vision
transformer from \citet{morenocartagena2025leveragingpretrainedvisualtransformers},
is referred to as the name of its base model “SWINv2” in text,
so we repeat that here. These results generally show our model as
a considerable improvement over ATAT and the SWINv2 model. ORACLE
was evaluated on \elasticc v2 with a smaller number of classes, but
the differences between \elasticc v1 and v2 are unclear, due to lack
of published articles on the matter. The ORACLE team makes comparisons
against ATAT, suggesting they believe \elasticc v1 and v2 numbers
are comparable, but a more 1-1 comparison would be strongly preferred.
Unfortunately, \elasticc preprocessing is tricky. While ORACLE has
also shared their preprocessing routines, the lack of standardization
means we could not immediately integrate it. We have made some suggestions
for improving this situation, which would facilitate more accurate
model comparisons and lead to better scientific understandings (see
Section \ref{subsec:FW-Better-standardization}). For model sizes:
ATAT's model sizes were not reported, but we loaded a checkpoint and
computed the total number of parameters for all tensors. The SWINv2
seems to be based on the “Tiny” variant; the v2 paper is not
open access, but the v1 paper lists this variant as having 29M parameters
\citet{Liu_2021}.

In Table \ref{tab:Fine-class-results}, we show fine-class performance
of ATCAT. Please see \Figref{Confusion-matrices-fine} for confusion
matrices. \citet{Shah_2025} notes that ATAT \citep{Cabrera-Vives_2024}
underperforms on CARTs compared to the random forest, but our model
has reversed that weakness and outperforms both the RF (random forest)
and ATAT on CART, although it is still one of the more difficult classes.
Please see \citet{Shah_2025} for a valuable discussion on CART classification.
Our model achieves leading F1 scores on all fine classes, but sometimes
makes a different precision-recall tradeoff. If a different precision-recall
tradeoff is desired, e.g. when selecting a particular class via a
“quality cut”, downstream users can choose different classifier
thresholds, or re-train our models using our Soft F$\beta$ loss for
different values of $\beta$.

\subsection{Unsupervised pretraining}\label{subsec:Results-pretraining}

\begin{table}
\begin{centering}
\begin{tabular}{|c|c|c|}
\hline 
Experiment & LC-only & LC+metadata\tabularnewline
\hline 
\hline 
Baseline$^{\text{(a)}}$ & \res{pretr true baseline lc_only f1}\% & \res{pretr true baseline lc_meta f1}\%\tabularnewline
\hline 
Increased regularization$^{\text{(b)}}$ & \res{pretr no ens lc_only f1}\% & \res{pretr no ens lc_meta f1}\%\tabularnewline
\hline 
Ensemble only$^{\text{(c)}}$ & \res{pretr ens lc_only f1}\% & \res{pretr ens lc_meta f1}\%\tabularnewline
\hline 
Pretrained ensemble$^{\text{(d)}}$ & \res{pretr setup A lc_only f1}\% & \res{pretr setup A lc_meta f1}\%\tabularnewline
\hline 
\end{tabular}
\par\end{centering}
\caption{\textbf{Unsupervised pretraining experiments}. We remove labels from
90\% of our normal training dataset, sampled uniformly at random.
These experiments build on each other. Row (a) is our model with no
modifications, trained on the remaining 10\% of training data with
labels. Row (b) increases regularization. Row (c) ensembles 10 sub-models.
Row (d) adds unsupervised pretraining on the full train dataset (unsupervised
pretraining does not use labels) and then does fine-tuning on the
remaining labeled 10\%. Values are Macro F1 scores, with $\pm$ from
\texttt{np.std} over 3 CV folds. See text for details.}\label{tab:Semi-Supervised}
\end{table}

\begin{figure}
\begin{centering}
\includegraphics[width=8cm]{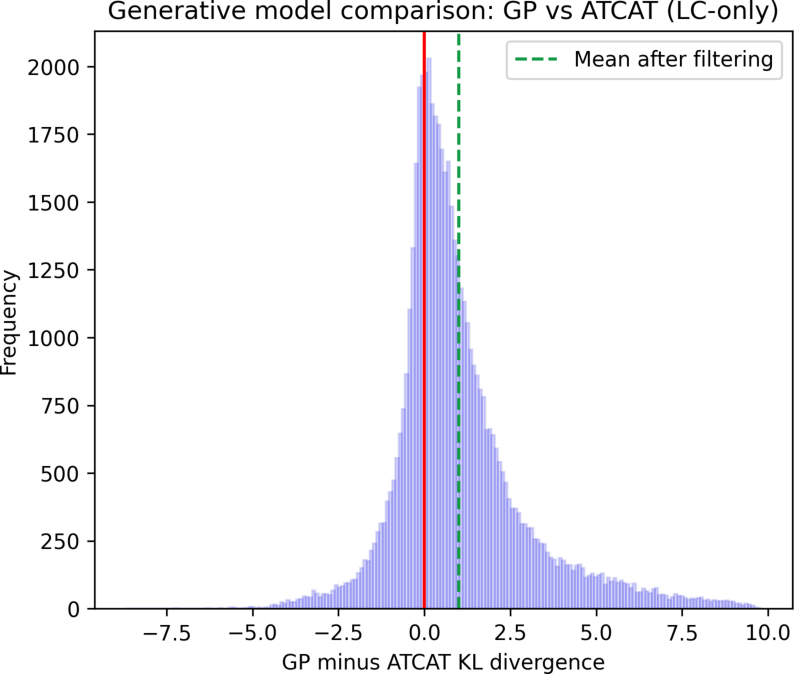}
\par\end{centering}
\caption{\textbf{Evaluation of generative modeling performance}. We predict
next LC points for a sample of 1000 light curves (50 per class) on
the test dataset. For each point, we compute the difference of forward
KL divergence; assuming that the model's flux is $P=\mathcal{N}\left(\text{flux},\text{flux\_err}\right)$
for the point in question, we then look at the predicted distribution
of flux $Q$ by both of these models, and then compute $\text{KL}\left(P||Q_{\text{GP}}\right)-\text{KL}\left(P||Q_{\text{ATCAT}}\right)$.
Remember that KL divergence is always non-negative and smaller values
are better; when the KL is zero then the model matches the observation
perfectly. Hence, when this difference is positive (right of the solid
red line), our model is better, and when it is negative, the GP is
better. See text for details.}\label{fig:generative-performance}
\end{figure}
As described in Section \ref{subsec:Methods-unsupervised-pretraining},
we measure the generative performance of our model compared with the
Gaussian Process of \citet{Boone_2019}. We first analyzed the extreme
values, those with KL divergence greater than 10. For our model, there
were 464 such LC points in 10 examples. Almost all of these (98\%)
had flux values more than 500, so we assume they are most related
to the known modeling deficiencies we mentioned in Section \ref{subsec:Methods-unsupervised-pretraining}.
For the GP model, there were 5968 such LC points in 862 examples.
Unlike ATCAT, about 90\% of the GP outliers had flux less than 500
(and the GP has no theoretical limitation on larger fluxes). From
manual inspection, these generally corresponded to higher flux values
and points at the beginning of the sequence.

We then filtered these points out, and focused on the common case,
graphing its performance in \Figref{generative-performance}. Our
model is significantly better; of these remaining 54,713 points, the
mean of $\text{KL}\left(P||Q_{\text{GP}}\right)-\text{KL}\left(P||Q_{\text{ATCAT}}\right)$
was 1.0. This outcome is consistent with our expectations, since the
transformer is trained on 1.5M light curves, whereas the GP just has
a few hyperparameters. However, \citet{Boone_2019} was the winner
of the PLASTICC Kaggle competition, hence we believe their GP's kernel
selection and hyperparameters have been well-tuned for this task,
and should be a reasonable baseline. While we used our LC-only model
for fair comparison to the GP (which does not access metadata), it
is interesting and potentially useful that we can condition our generation
on the metadata.

We then look classification performance when removing labels from
90\% of our training split, in Table \ref{tab:Semi-Supervised}. The
simple step of increasing regularization is critical, as shown in
the second row (settings in Section \ref{subsec:Methods-training-routine}).
Ensembling was also quite helpful; we hypothesize that it functions
similar to regularization, and ensures that no single example can
disproportionately influence model output. We ensembled by splitting
our dataset into 5 sub-folds (having already applied the main cross-validation
split) and training 10 models, each with one sub-fold omitted. Our
pretraining step provided less than 1\% absolute F1 gain on both LC-only
and LC+metadata, but at the scale of LSST we believe it is valuable.
Furthermore, as argued in the introduction, we believe that this scenario
will correspond to practical situations of having a vast quantity
of unlabeled data (even more than 10:1), and pretrained models should
have better generalization and more useful last-layer embeddings.

\subsection{Early detection}

\begin{figure*}
\begin{centering}
\includegraphics[width=16.5cm]{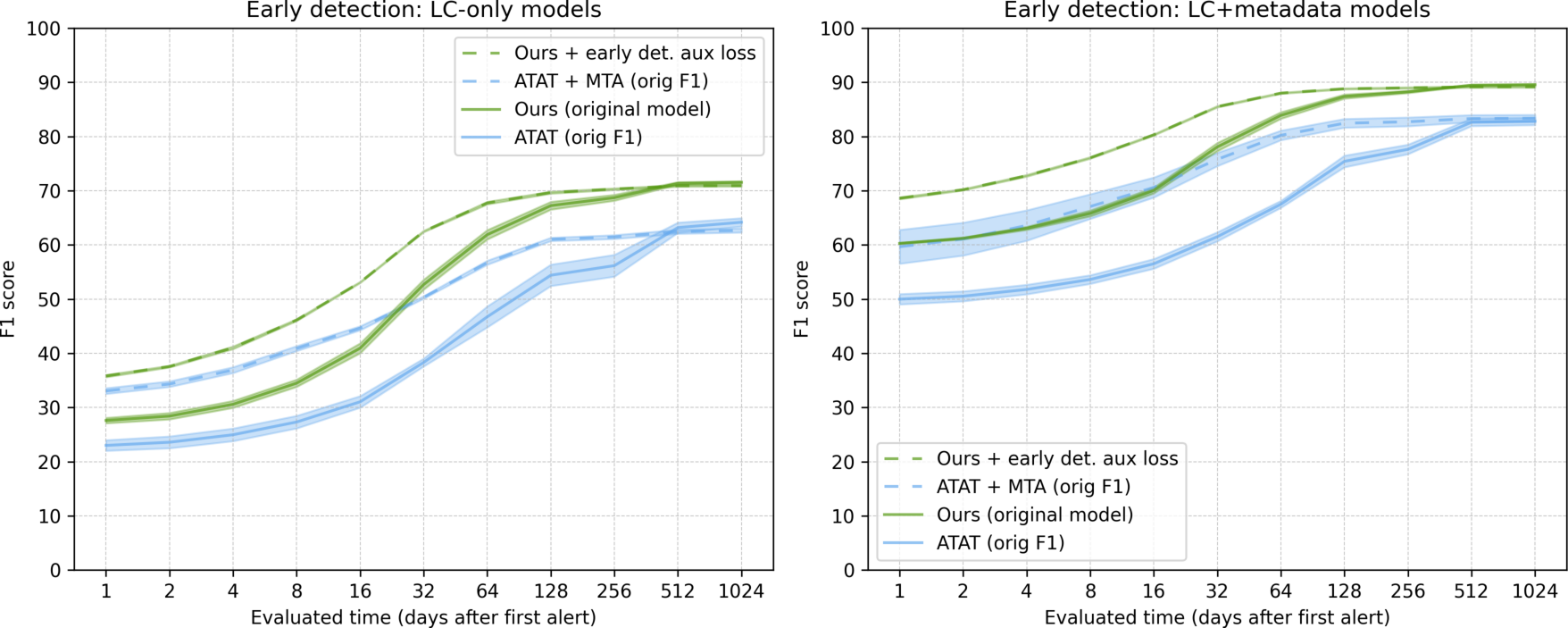}
\par\end{centering}
\caption{\textbf{Early detection performance}. The strong performance of ATCAT
carries over to early detection setups. Variances are \texttt{np.std()}
calculations over the 5 CV folds for both ATAT and ATCAT. See text
for details.}\label{fig:Results-early-det}

\end{figure*}

In \Figref{Results-early-det}, we show early detection performance.
When we added an early detection auxiliary loss described in Section
\ref{subsec:Methods-Early-detection} (dashed green line), both our
LC-only and LC+metadata models outperform ATAT variants at all time
scales, including ATAT's MTA variant (masked temporal augmentation,
a similar strategy to improve early detection performance, which uses
data augmentation to truncate light curves at specific time cutoffs).
Our model's variance across CV folds was also much more tightly controlled,
which is generally expected as model quality improves. However, we
noticed slight drops in final accuracy when adding the early detection
loss, suggesting some remaining headroom in our model (see Section
\ref{subsec:FW-modeling}).

\subsection{Calibration}

\begin{figure}
\begin{centering}
\includegraphics[width=8cm]{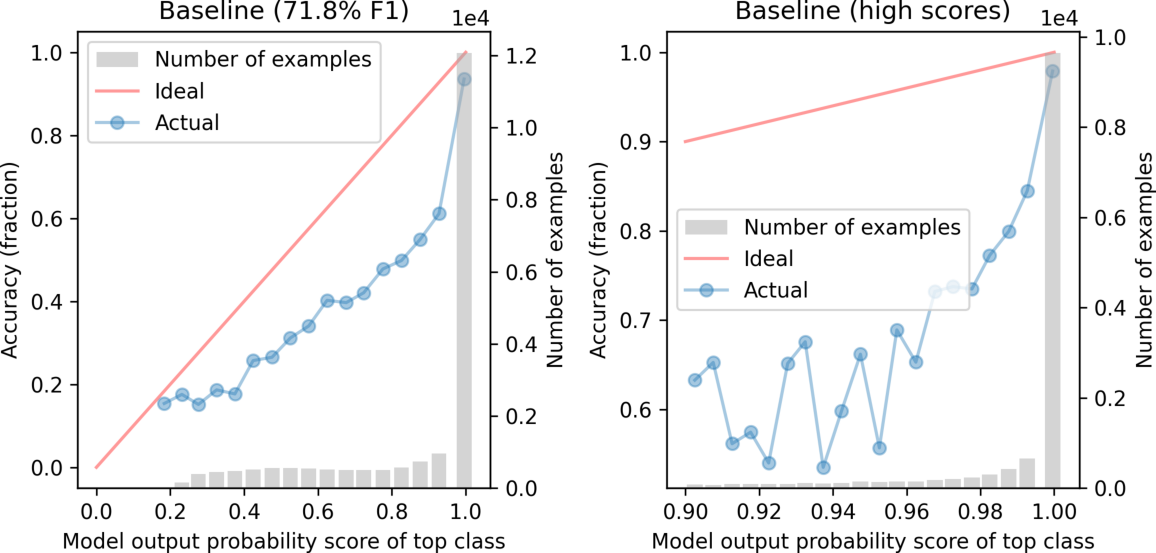}
\par\end{centering}
\vspace{0.3cm}
\begin{centering}
\includegraphics[width=8cm]{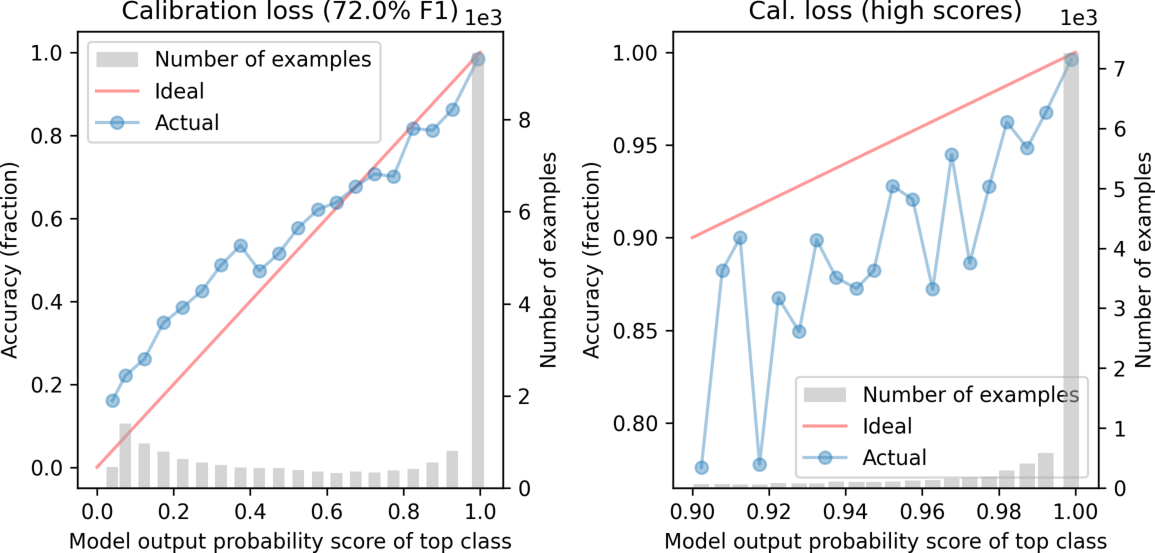}
\par\end{centering}
\vspace{0.3cm}
\begin{centering}
\includegraphics[width=8cm]{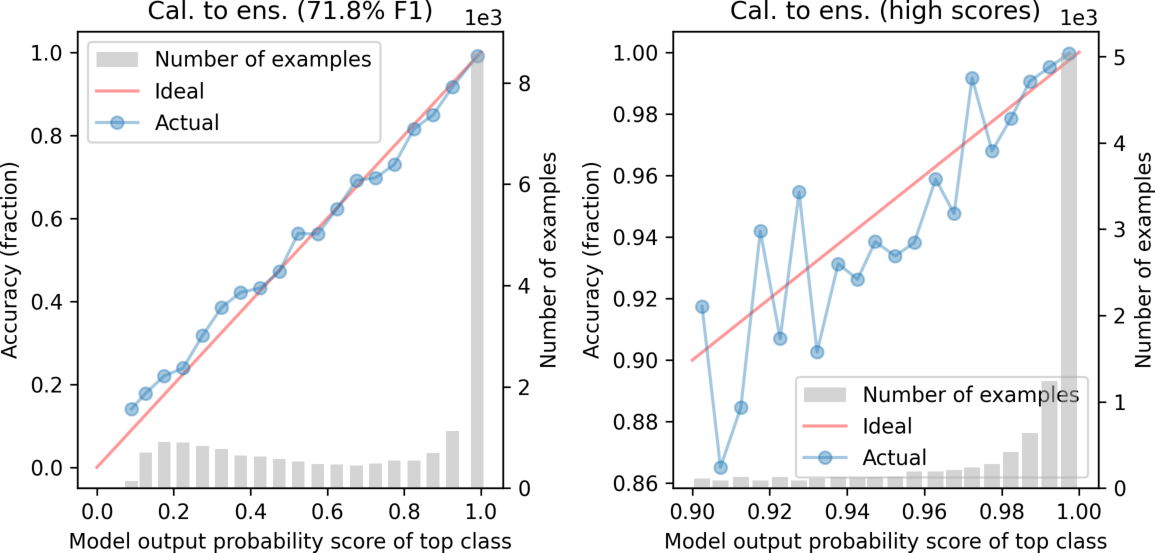}
\par\end{centering}
\caption{\textbf{Calibration experiments}. All models here are LC-only, trained
on the first CV split; their performance on the test set at step 36000
is shown above. Each plot is a reliability diagram, which correlates
the “confidence” of a model with its actual accuracy. Values
above the red line indicate that the model is underconfident; values
below indicate that the model is overconfident / “confidently wrong”.
In this case of multi-class prediction, we interpret the “confidence”
to be the probability of the top class (see text) and per-example
“accuracy” to be 1 if the top predicted class (which can be
less than 0.5) is the same as the actual class (ignoring the “unsure”
output class) and 0 otherwise. The number of examples in each confidence
bin is shown with grey bars. Each row is for a single model, for the
entire range of output scores, and zoomed to the more confident ones.}\label{fig:calibration-results}
\end{figure}

Results for our calibration experiments for LC-only models can be
found in \Figref{calibration-results}. Following standard practice
in \citet{Guo_2017}, we used the score of the top-scoring class as
the confidence, which both worked well in practice and works for models
which are not trained to output good $p_{\text{unsure}}$ scores.
Models are free to generate confidence scores through any method though.
After binning by model confidence, we adjusted the x-axis value to
be the actual mean of binned samples; we believe this creates slightly
more accurate charts when comparing to the red “ideal” line,
although bins/points are not exactly evenly spaced. Results from LC+metadata
models were similar, and are omitted for space reasons.

Our top-F1-scoring model is poorly calibrated. For example, when it
outputs a probability score of “94\%” for a class, it is only
right 60\% of the time. The training-time calibration loss is much
better, and the bootstrap ensemble is extremely good, albeit being
twice as slow to train. 

Label smoothing (following \citealp{Szegedy_2015}) produced model
outputs that were closer to the calibration line, but in a way that
very much reflected its mechanics: most top scores were now around
0.95 and few were highly confident. We believe this approach is less
useful than the others; after all, one could take a baseline model
and shift over all of its values, but it fails to further separate
examples into ones the model is actually confident in. As we can see
in our results, successful calibration techniques do still output
high scores frequently for the \elasticc dataset.

Focal loss with $\gamma=0.5$ ended up producing a very well-calibrated
model, but with a slightly greater cost to F1 (71.3\%). It has the
advantage of being a very simple technique, but higher values of $\gamma$
end up with an under-confident model, necessitating tuning. We omit
charts for space reasons.

\subsection{Ablation studies}

\subsubsection{Input encoder}\label{subsec:Input-encoder-ablation}

\begin{table}
\begin{centering}
\begin{tabular}{|c|c|c|}
\hline 
Experiment & LC-only & LC+metadata\tabularnewline
\hline 
\hline 
ATAT embedder$^{\text{(a)}}$ & \res{abl emb lc_only legacy_atat}\% & \res{abl emb lc_meta legacy_atat}\%\tabularnewline
\hline 
ATAT + tanh inp. flux$^{\text{(b)}}$ & \res{abl emb lc_only legacy_atat_with_tanh}\% & \res{abl emb lc_meta legacy_atat_with_tanh}\%\tabularnewline
\hline 
\begin{cellvarwidth}[t]
\centering
Posenc, no ch, no

flux err, no scaling$^{\text{(c)}}$
\end{cellvarwidth} & \res{abl emb lc_only pre_rotary_no_scaling_no_ch_no_flux_err}\% & \res{abl emb lc_meta pre_rotary_no_scaling_no_ch_no_flux_err}\%\tabularnewline
\hline 
\begin{cellvarwidth}[t]
\centering
Posenc, no flux err,

no scaling$^{\text{(d)}}$
\end{cellvarwidth} & \res{abl emb lc_only pre_rotary_no_scaling_no_flux_err}\% & \res{abl emb lc_meta pre_rotary_no_scaling_no_flux_err}\%\tabularnewline
\hline 
Posenc, no scaling$^{\text{(e)}}$ & \res{abl emb lc_only pre_rotary_no_scaling}\% & \res{abl emb lc_meta pre_rotary_no_scaling}\%\tabularnewline
\hline 
Posenc$^{\text{(f)}}$ & \res{abl emb lc_only pre_rotary}\% & \res{abl emb lc_meta pre_rotary}\%\tabularnewline
\hline 
\begin{cellvarwidth}[t]
\centering
\textbf{ATCAT default}$^{\text{(g)}}$

(rotary encoder,

wavelengths)
\end{cellvarwidth} & \res{abl emb lc_only current_rotary}\% & \res{abl emb lc_meta current_rotary}\%\tabularnewline
\hline 
\end{tabular}
\par\end{centering}
\caption{\textbf{Ablations for embedding schemes}. We break down our improvements
over ATAT's embedder into a variety of sub-steps. Our embedding scheme
is the most significant accuracy gain in this work. All rows use the
rest of our model/training regime, modifying only the embedder. Rows
(a) and (b) use the ATAT embedder. Row (c) switches to a variant of
\citet{Vaswani_2017}, although it chooses a set of time value scales
appropriate to \elasticc. In row (d) we add channel index information
through an indexed linear embedding, in row (e) we add \texttt{flux\_err},
in row (f) we add some dynamic range pre-scaling. In row (g) we replace
the indexing channel approach from row (d) with our rotary encoder
based approach, making the model more generally useful for non-\elasticc
data. Values are Macro F1 scores, with $\pm$ from \texttt{np.std}
over 3 CV folds. See text for details.}\label{tab:Abl-Embeddings}
\end{table}

In \Tabref{Abl-Embeddings} we examine varying input encoder changes.
Our embedding scheme is the most significant accuracy gain in this
work. Our default/final model is the last row, described in Section
\ref{subsec:Methods-input-encoding}. In rows (a) and (b), we use
our model with the ATAT embedder (slightly re-implemented to accommodate
appropriately shorter sequence lengths, as discussed in Section \ref{subsec:Results-comp-perf};
we wrote tests to ensure equivalence). These results do not match
the ATAT numbers exactly because we did not adjust other parts of
the model; in particular, the LC+metadata models did not use the dynamic
feature embedder, which includes many features based on light curve
flux values. In row (b), we replacing calibrated \texttt{flux} with
\texttt{tanh(flux/10)} (dividing by 10 to preserve dynamic range).
We saw an improvement on LC+metadata but unexpected decrease on LC-only.

In row (c), we use an embedder which is closer to \citet{Vaswani_2017},
except with the important scaling of time values (see Section \ref{subsec:Methods-input-encoding})
to accommodate the uneven distribution we have, and tanh scaling of
input flux at varying scales. Please see \citet{Cabrera-Vives_2024}
for a more literal implementation of \citet{Vaswani_2017}'s embedder.
In row (d), we add the channel information, in this case performing
part of the linear transformation of time values indexed by channel.
The ATAT experimentation with \citet{Vaswani_2017}'s positional encoder
did not include channel information---while this is not part its
definition, it is an additional piece of data which is quite impactful.
In row (e), we add the \texttt{flux\_err} information, which provides
a significant boost. In row (f), we add the scaling discussed in Section
\ref{subsec:Methods-input-encoding}, which we implemented after comparing
the dynamic range of activations when debugging our network; despite
seeming arbitrary, it appears to be helpful. Finally, in row (g),
we switch to a rotary encode using channel wavelengths. This provides
a small accuracy boost for LC+metadata and less for LC-only. More
importantly, our model more widely useable, since it is no longer
tied to LSST color bands.\footnote{As discussed in Methods, we do normalize the range by LSST bands,
but this could be relaxed, and the model does not fail for wavelengths
out of that range (which are created by our data augmentation process).}

\subsubsection{Data augmentation}\label{subsec:Data-aug-results}

\begin{table}
\begin{centering}
\begin{tabular}{|c|c|c|c|}
\hline 
Experiment & Rate & LC-only & LC+metadata\tabularnewline
\hline 
\hline 
\begin{cellvarwidth}[t]
\centering
\textbf{Baseline}

No aug.
\end{cellvarwidth} &  & \res{abl data aug lc_only baseline}\% & \res{abl data aug lc_meta baseline}\%\tabularnewline
\hline 
Flux scaling & 0.2 & \res{abl data aug lc_only only flux scale}\% & \res{abl data aug lc_meta only flux scale}\%\tabularnewline
\hline 
Redshifting & 0.1 & \res{abl data aug lc_only only redshift}\% & \res{abl data aug lc_meta only redshift}\%\tabularnewline
\hline 
Subsampling & 0.25 & \res{abl data aug lc_only only subsample}\% & \res{abl data aug lc_meta only subsample}\%\tabularnewline
\hline 
Time shifting & 0.2 & \res{abl data aug lc_only only time shift}\% & \res{abl data aug lc_meta only time shift}\%\tabularnewline
\hline 
\textbf{ATCAT default} &  & \res{abl data aug lc_only scheme1}\% & \res{abl data aug lc_meta scheme1}\%\tabularnewline
\hline 
Random noise & 0.15 & \res{abl data aug lc_only only random noise}\% & \res{abl data aug lc_meta only random noise}\%\tabularnewline
\hline 
Preset 2 & {*} & \res{abl data aug lc_only scheme3}\% & \res{abl data aug lc_meta scheme3}\%\tabularnewline
\hline 
\end{tabular}
\par\end{centering}
\caption{\textbf{Ablations for data augmentation}. We examine the effect of
adding various data augmentations. Each row adds to the baseline;
rows are not iteratively stacked, although the “ATCAT default”
preset applies the 4 above augmentations, and “Preset 2” adds
additional random noise to the ATCAT default. “Preset 2” was
used for pretraining experiments, where we had less labeled data and
wanted to add more regularization. Rates were lightly tuned. Values
are Macro F1 scores from a single experiment on CV fold 0, so please
beware only larger differences should be relevant. See text for details.}\label{tab:Abl-Data-Aug}
\end{table}
We implemented several types of data augmentation described in Section
\ref{subsec:Data-aug-methods}. Our results for these appear in Table
\ref{tab:Abl-Data-Aug}. The subsampling data augmentation routine
is the most effective. We hypothesize that this augmentation could
be particularly effective because it always results in another valid
light curve from that class.

\subsubsection{Model architecture}\label{subsec:model-arch-ablation}

\begin{table}
\begin{centering}
\begin{tabular}{|c|c|c|}
\hline 
Experiment & LC-only & LC+metadata\tabularnewline
\hline 
\hline 
\textbf{ATCAT default}$^{\text{(a)}}$ & \res{abl model lc_only baseline}\% & \res{abl model lc_meta baseline}\%\tabularnewline
\hline 
Larger model$^{\text{(b)}}$ & \res{abl model lc_only larger}\% & \res{abl model lc_meta larger}\%\tabularnewline
\hline 
Non-hybrid model$^{\text{(c)}}$ & \res{abl model lc_only non_hybrid}\% & \res{abl model lc_meta non_hybrid}\%\tabularnewline
\hline 
\end{tabular}
\par\end{centering}
\caption{\textbf{Ablations for non-embedding architecture choices}. We explore
two separate model variants. In row (b) we moderately increase several
hyperparameters affecting model size. In row (c) we replace the local/hybrid
attention with only global attention layers. Values are Macro F1 scores,
with $\pm$ from \texttt{np.std} over 3 CV folds. See text for details.}\label{tab:Abl-Model}
\end{table}

In \Tabref{Abl-Model} we perform ablations on a few modeling choices
(see section above for the more impactful input embedding choices).
We experimented with a larger model with 512 embedding dimension (instead
of 384), 8 layers (instead of 4), and 128 attention dimension (shared
among 4 attention heads, so 32 per attention head). This was significantly
slower and not much better, but may be worth trying on different datasets.
Earlier experiments with local attention showed a small benefit, but
in \Tabref{Abl-Model} we did not see much impact, and the “non-hybrid
model” with only global attention layers worked just as well. However,
if combined with a limit on the number of points which can attend
to each other (say at most $k$ points, even if we get an outlier
cluster with more than $k$ points within 1 day / 10 days), then local
attention layers can be $\mathcal{O}\left(kn\right)$ rather than
$\mathcal{O}\left(n^{2}\right)$, so this architecture may still be
advantageous for speed even if it does not provide an accuracy boost.

\subsubsection{Losses and optimization}\label{subsec:results-loss-opt}

\begin{table}
\begin{centering}
\begin{tabular}{|c|c|c|}
\hline 
Experiment & LC-only & LC+metadata\tabularnewline
\hline 
\hline 
\textbf{ATCAT default} & \res{abl loss opt lc_only baseline}\% & \res{abl loss opt lc_meta baseline}\%\tabularnewline
\hline 
Log-loss only & \res{abl loss opt lc_only logloss_only}\% & \res{abl loss opt lc_meta logloss_only}\%\tabularnewline
\hline 
Soft F1 loss only & \res{abl loss opt lc_only f1_only}\% & \res{abl loss opt lc_meta f1_only}\%\tabularnewline
\hline 
No “unsure” output & \res{abl loss opt lc_only no_extra_output_class}\% & \res{abl loss opt lc_meta no_extra_output_class}\%\tabularnewline
\hline 
ADAM, no LR sched. & \res{abl loss opt lc_only default_adam_no_sched}\% & \res{abl loss opt lc_meta default_adam_no_sched}\%\tabularnewline
\hline 
\end{tabular}
\par\end{centering}
\caption{\textbf{Ablations for loss functions}. We explore variants of losses
and optimizer settings. Experiments should be interpreted as modifying
the ATCAT default, and are not stacked on each other. Values are Macro
F1 scores, with $\pm$ from \texttt{np.std} over 3 CV folds. See text
for details.}\label{tab:Abl-Losses}
\end{table}

We also experimented with a few variants of loss functions and optimization
methods in \Tabref{Abl-Losses}. In general, the log-loss is essential
(i.e. only using Soft F1 loss underperforms significantly). Our default
model, which adds Soft F1, does seem to provide a barely-significant
performance boost. We also did an ablation, changing from our somewhat
uncommon optimizer setting with $\beta_{1}=0$ (following ATAT) to
a more vanilla configuration, although maintaining the 1e-4 learning
rate. This was significantly worse for the LC+metadata experiments.

\subsection{Computational performance}\label{subsec:Results-comp-perf}

\begin{table}
\begin{centering}
\begin{tabular}{|c|c|c|}
\hline 
Experiment & CPU/GPU & Rate {[}LCs/sec{]}\tabularnewline
\hline 
\hline 
ATAT feature extraction & 1c CPU & \res{perfbench atat feat extraction}\tabularnewline
\hline 
\emph{above, scaled to approx. multicore} & {*} & \res{perfbench atat feat extraction scaled}\tabularnewline
\hline 
ATAT model without features & A100 & \res{perfbench atat model no feats}\tabularnewline
\hline 
ATAT model with features & A100 & \res{perfbench atat model with feats}\tabularnewline
\hline 
Our model, LC only & 40GB A100 & \res{perfbench our model lc only 40gb}\tabularnewline
\hline 
Our model, LC only & 80GB A100 & \res{perfbench our model lc only 80gb}\tabularnewline
\hline 
Our model, LC + Meta & 40GB A100 & \res{perfbench our model lc meta 40gb}\tabularnewline
\hline 
Our model, LC + Meta & 80GB A100 & \res{perfbench our model lc meta 80gb}\tabularnewline
\hline 
ORACLE & H100 & \res{perfbench oracle}\tabularnewline
\hline 
Our model, LC + Meta & H100 & \res{perfbench our model lc meta h100}\tabularnewline
\hline 
Our model, LC + Meta & RTX 4090 & \res{perfbench our model lc meta rtx 4090}\tabularnewline
\hline 
\end{tabular}
\par\end{centering}
\caption{\textbf{Inference performance}. We measure the inference performance
of ATCAT, our model. ATAT feature extraction was reported as a single-core
number, so we have scaled it to approximate performance-per-watt of
the A100 (see Appendix \ref{subsec:multicore-CPU-scale} for details).}\label{tab:Inference-Perf}
\end{table}

We achieved generally favorable GPU performance over other state-of-the-art
\elasticc classification models. First, we used PyTorch compilation,
which does a variety of optimizations and almost always results in
significant speedups. Secondly, we matched the maximum sequence length
of \elasticc light curves exactly (243 points). In comparison to
ATAT specifically, their model pads all channels to the same length,
resulting in a sequence length of 384. Since the attention mechanism
is $\mathcal{O}\left(n^{2}\right)$, this means our model only needs
40\% of the computation for this step. Of course, this advantage will
disappear if we need to increase sequence length to support longer
non-simulated light curves. Finally, we used \texttt{bfloat16} pervasively
after input encoding. Half-precision floating point values are much
faster on modern GPUs; the A100 processes \texttt{float32} data at
19.5 Tflops, but \texttt{bfloat16} data at 312 Tflops (16x faster).
The difference has only widened on more modern chips such as the H200,
which processes \texttt{float32} data at 60 Tflops and \texttt{bfloat16}
data at 1979 Tflops (30x faster).

Our results are in \Tabref{Inference-Perf}. We threw out the first
100 batches to ignore compilation and dataset prefetching. We report
ATAT and ORACLE \citep{Shah_2025} results for batch size 2000, and
our results at batch size 2048 for the A100 and 4096 for the H100.
While larger batches sizes are important, around these optimal values,
changes by a factor of 2 or 4 were not important. We also report RTX
4090 results (a consumer GPU 1/8 to 1/3 the cost of the A100), and
recommend this setup.

Versus ATAT, the largest benefit is that our model does not require
/ benefit from CPU-bound feature extraction. ATAT's best-performing
model includes dynamic features, which need to be extracted every
time for optimal performance. In other words, unlike most metadata,
they cannot be cached for one object and re-used as new LC measurements
come in. After applying a reasonable scaling to ATAT's single-core
results (see Appendix), we find that our model is approximately \res{our model times faster vs atat}$\times$
faster. ATAT also has a model variant that skips these dynamic features
for a modest accuracy penalty; compared to this setup, our model is
\res{our model times faster vs atat gpu model only}$\times$ faster.
We ran head-to-head comparisons with ORACLE as well, a RNN-based network
which is focused more heavily on high throughput. On the H100, our
model is \res{our model times faster vs oracle}$\times$ faster.
However, RNNs' sequential nature generally means they perform worse
on GPUs and relatively better on CPUs, so we expect that ORACLE has
better CPU performance. We haven't had time to investigate CPU performance
of ATCAT; please see Section \ref{subsec:FW-throughput-perf}.

\subsubsection{Training performance}

We did light optimization work for training speed. This was helpful
for development speed and costs. Our data loader, with all augmentations,
can load light curves at 192 batches/sec on a AMD Ryzen 9 7900X (batch
size 256, so 49,000 LC/s). We achieved this while using a Polars dataframe-based
implementation, which allowed for expressive runtime transformations.
The trainer usually runs around 35 steps/s on a RTX 4090 (batch size
256, so 9,000 LC/s). Full training runs usually take around 45 minutes
on an RTX 4090 and cost around \$0.35 on our cloud host; by contrast,
\citet{Shah_2025} reports 12 hour training times and 4 hour evaluation
times on an H100 GPU, which is considerably more expensive.

\section{Discussion and further work}

In this section, we discuss the value of ATCAT, provide confusion
matrices, discuss some possible approaches for anomaly detection,
and share our top ideas for improvement.

\subsection{Summary and value of contributions}

In this work, we have shown that our model \nicknameWithEmoji
significantly advances the state of the art in \elasticc classification
accuracy. We have also shown that it can be tuned for label-efficient
training and generative modeling, early classification, and calibrated
classifications.

As real LSST data becomes available, we hope that our recipes for
unsupervised pretraining can be leveraged to make use of vast quantities
of unlabeled data, and to train effective classifiers when labeled
data is harder to come by. For example, for some tasks it may be desirable
to train models based on expert-labeled classifications, rather than
simulator outputs, and producing a high-quality labeled dataset as
large as \elasticc would likely be infeasible. Employing all of our
techniques in Section \ref{subsec:Results-pretraining} should be
a good first step. Domain adaptation is a challenging problem however,
and much more work will be needed, as we discuss later.

For early detection, we demonstrated state of the art results with
meaningful improvements. This performance boost will result in the
ability to follow up on more objects of interest, or reduce “wasted”
telescope time following up on less interesting objects. These follow-up
observations usually use a more narrow field of view telescope, and/or
telescopes with spectroscopic tools, which are in high demand for
a variety of science goals. 

For cosmological and stellar population studies, our work on accuracy-calibrated
versions of ATCAT should be useful, since selecting by confidence
score of a calibrated model allows selection by expected model precision.
Our improvements in accuracy should mean that if researchers use our
models for quality cuts, they will be able to select more objects
at the same level of precision. However, distributional shifts remain
a problem, and calibration does not mitigate classifier bias, as we
discuss later.

Finally, our basic work on inference performance should be quite impactful.
We hope that ATCAT can be run regularly as part of LSST alerting pipelines,
for larger surveys, and anything in-between. ATCAT's inference throughput
implies a theoretical lower bound cost of \$3 per billion objects
(33,000 LC/s on a RTX 4090 cloud worker at \$0.35/hour), though full
pipeline deployment will undoubtedly be more expensive. Faster models
also allow for faster experimentation cycles. Lastly, analysis of
LSST data is a task carried out by scientists throughout the world,
some of whom do not have supercomputer access. We have managed to
both significantly improve upon state-of-the-art accuracy, and create
a model that is faster and less expensive to run.

\subsection{Confusion matrix}

\begin{figure*}
\begin{centering}
\includegraphics[width=2\columnwidth]{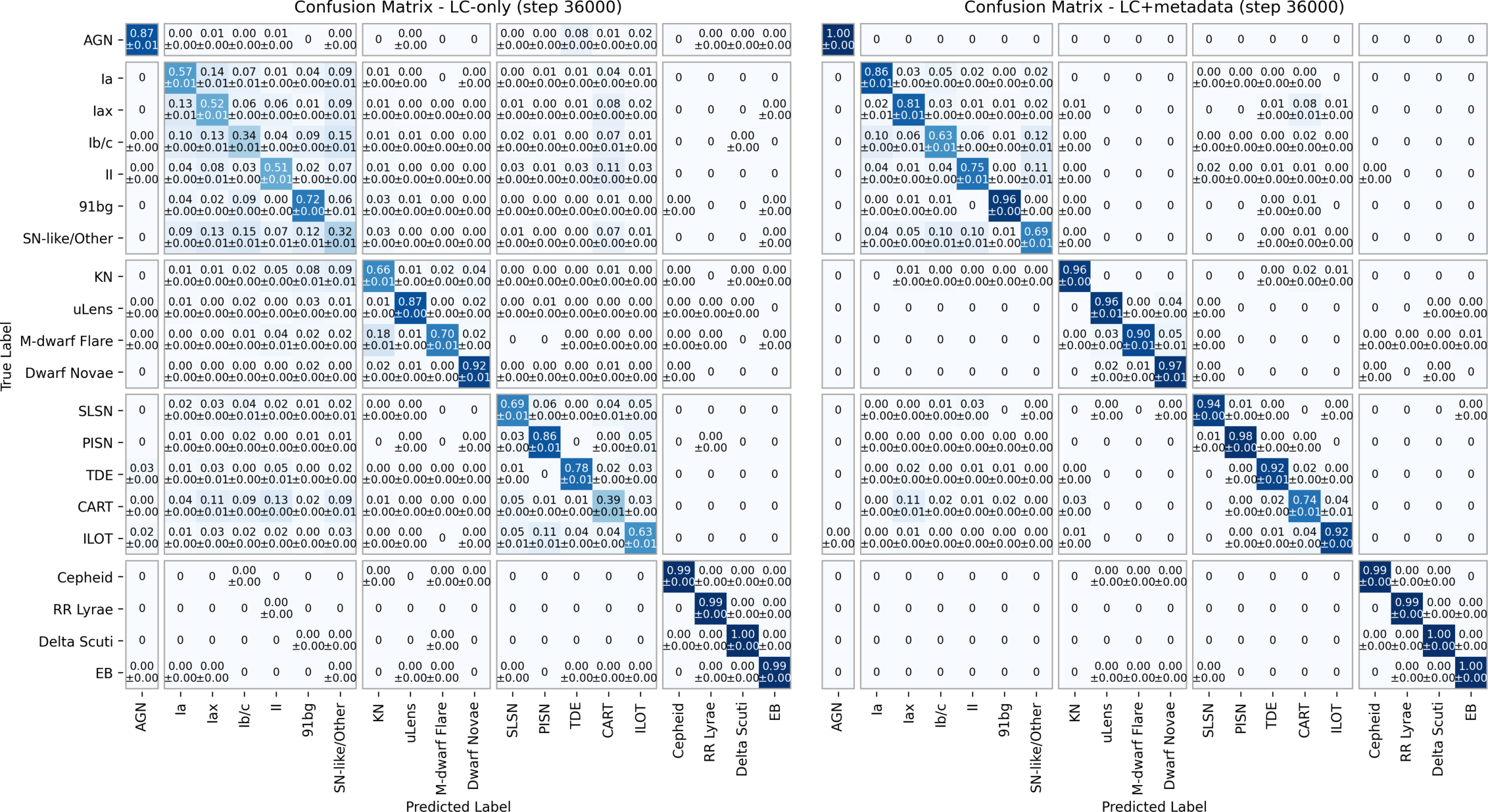}
\par\end{centering}
\caption{\textbf{Confusion matrices}. For each fold, we first accumulated
counts indexed by true label and predicted label. Then we normalized
by the true label, so each row sums to one. Finally, we take the mean
and \texttt{np.std} over all of these matrices and visualize the result.
We have segmented the labels into classes similar to \citet{Shah_2025},
so values within blocks on the diagonal can be interpreted as confusion
between fine classes of the same coarse class, whereas blocks off
the diagonal can be interpreted as confusion between coarse classes
(for some purposes, this may be a more serious error).}\label{fig:Confusion-matrices-fine}
\end{figure*}
We provide a confusion matrix for LC-only and LC+metadata variants
of ATCAT in \Figref{Confusion-matrices-fine}, for comparison with
other papers, and understanding which specific classes are confused
(beyond per-class F1 scores in Table \ref{tab:Fine-class-results}).
We grouped the confusion matrix by coarse class, which helps clarify
errors between fine classes and between coarse classes, and have published
this code for any subsequent work to utilize if they wish. We refer
readers to \citet{Shah_2025} for a good analysis / discussion of
some common confusions. Although ATCAT has increased accuracy, its
pairwise confusions remain similar to other works. Finally, despite
all of the variable stars (bottom right of the confusion matrix) having
high scores, we noticed that some of the misclassified ones could
be period folded and identified as their true class. Therefore, for
astronomers doing variable star studies, if maximum accuracy is critical,
we recommend combining ATCAT with period folding logic (possibly adding
the detected period(s) as input features).

\subsection{Preliminary work: Anomaly detection}\label{subsec:Preliminary-work-Anomaly}

\begin{figure*}
\begin{centering}
\includegraphics[width=6cm]{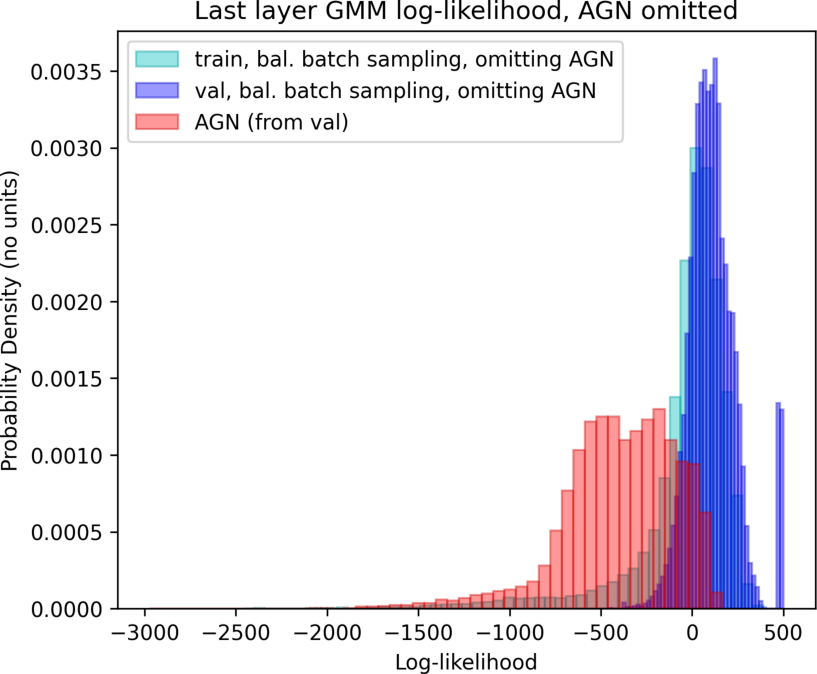}~~\includegraphics[width=6cm]{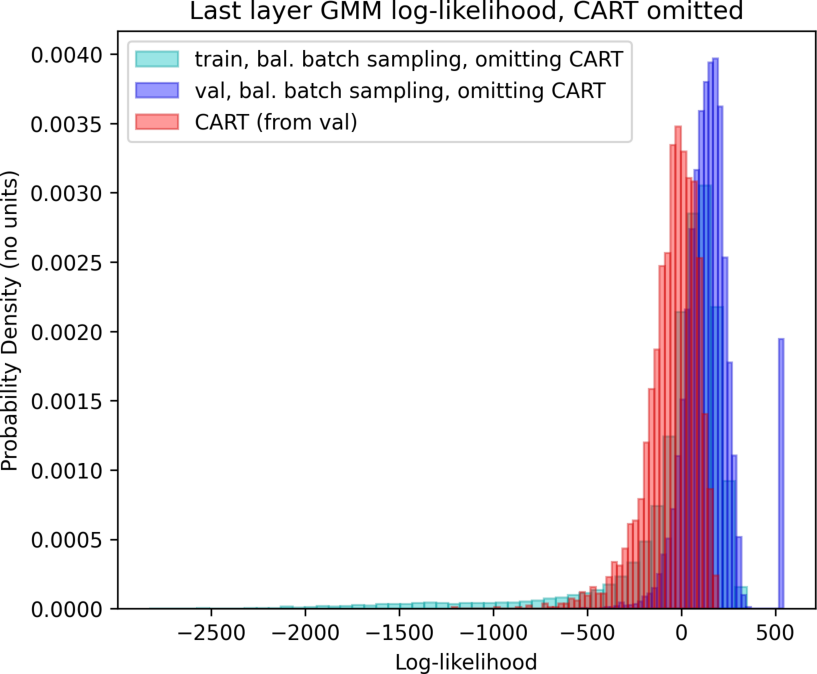}
\par\end{centering}
\caption{\textbf{GMM clustering for anomaly detection} (preliminary work).
We apply train a GMM on the last layer embeddings and see if it can
detect outliers. The GMM was trained on a balanced-batch version of
our validation set (the ATCAT model was trained on the training set,
minus the class in question). What we're looking for is separation
of our “novel” (held out) class from the rest of the data; see
text.}\label{fig:GMM-anomaly-det}
\end{figure*}

\begin{figure*}
\begin{centering}
\includegraphics[width=5.5cm]{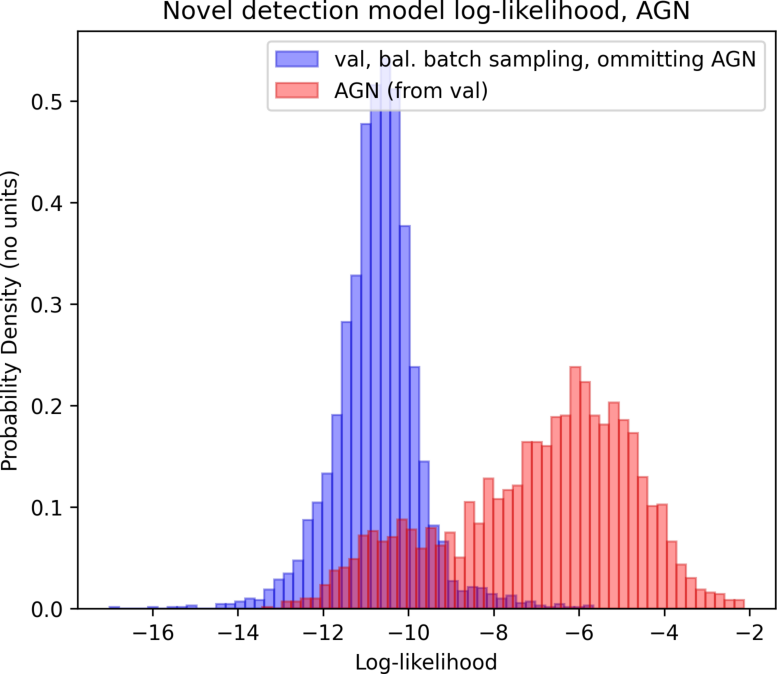}~~~\includegraphics[width=5.5cm]{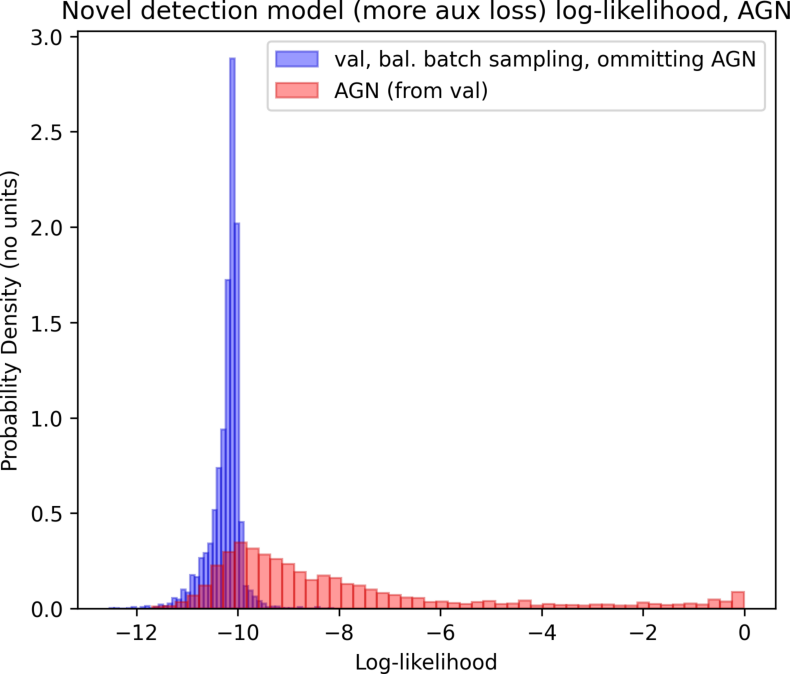}~\includegraphics[width=5.5cm]{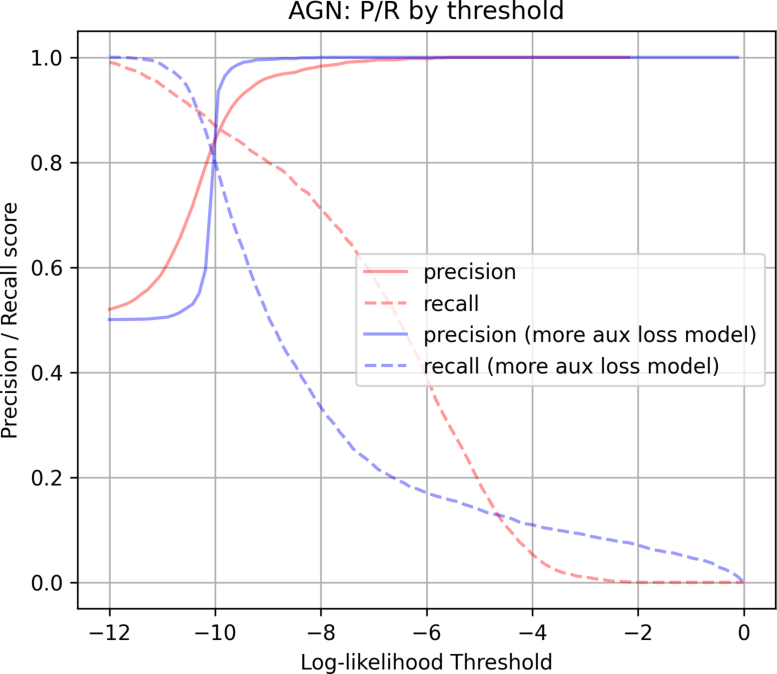}
\par\end{centering}
\vspace{0.3cm}
\begin{centering}
\includegraphics[width=5.5cm]{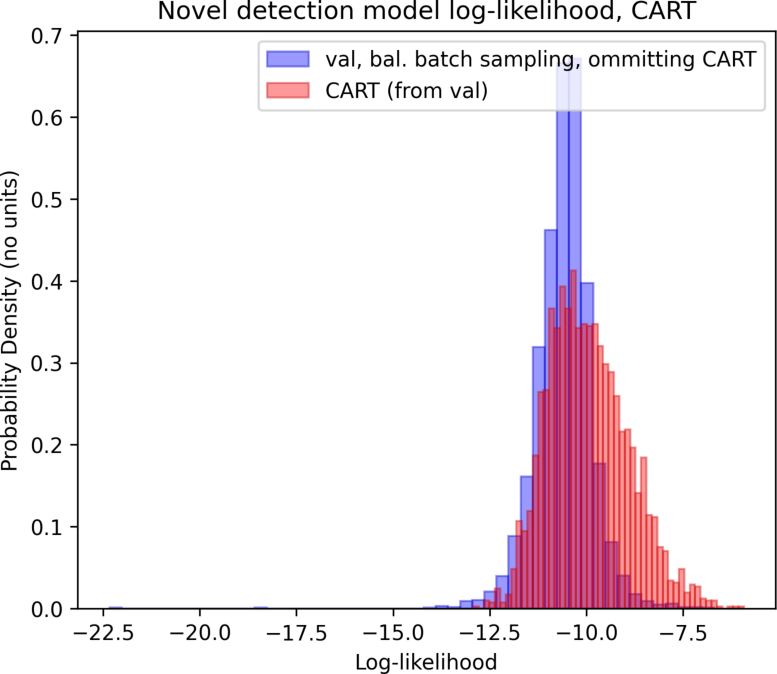}~~~\includegraphics[width=5.5cm]{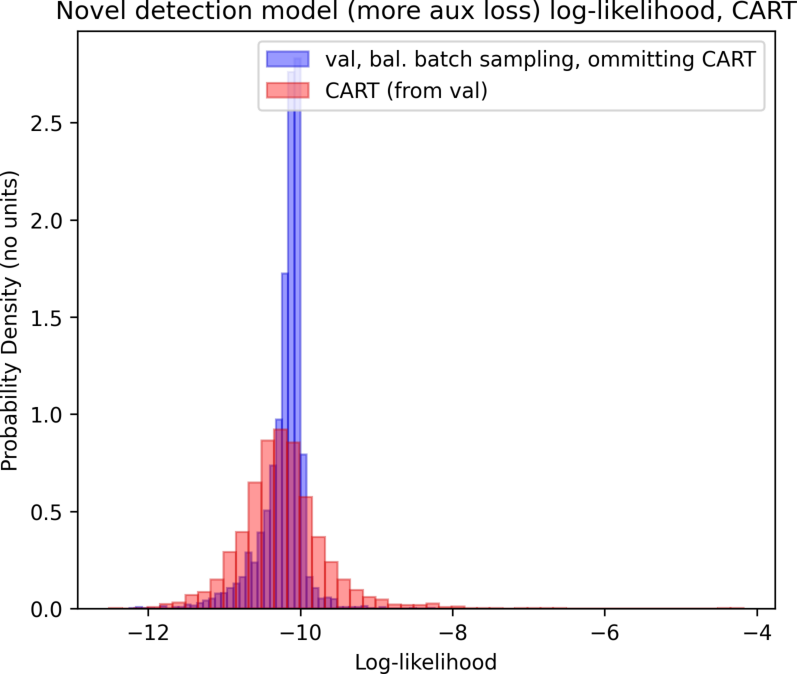}~\includegraphics[width=5.5cm]{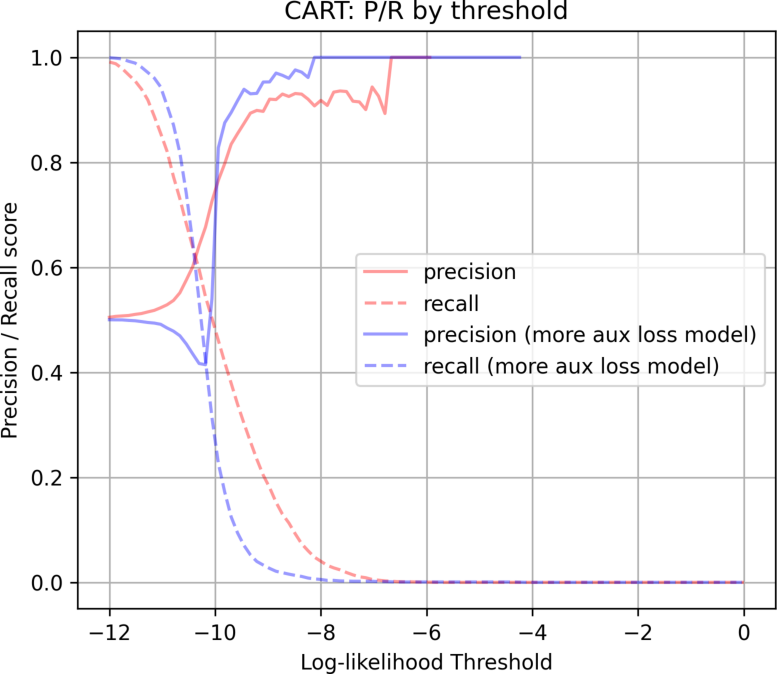}
\par\end{centering}
\caption{\textbf{Model variants for anomaly detection} (preliminary work).
We train a variant of ATCAT that has an extra “novel” output
class, and train the final classifier to output this class when last-layer
embeddings are anomalous. The left and center columns show novel detection
logit outputs. The first row shows results for AGNs (easy), the second
for CARTs (harder). The left column has a base weighting of the auxiliary
loss, and the middle column weights this loss even more, and further
shrinks random vector lengths to match the actual data standard deviations.
In these setups, train and test distributions (shown in \Figref{GMM-anomaly-det})
were very similar to val, so we omitted them for readability. What
we're looking for is separation of our “novel” (held out) class
from the rest of the data. The right column show precision/recall
by threshold, if we were to retrieve all examples greater than a certain
score threshold.}\label{fig:Novel-det-model}
\end{figure*}

We perform some preliminary investigation into the ability to use
ATCAT embeddings to detect novel types of objects. Work into anomaly
detection has traditionally used isolation forests, and custom algorithms
have also been applied \citep{Ishida_2019}. Our experimental setup
consists of training ATCAT with one or more classes of objects held
out, and then investigating the embeddings of those held-out object
classes. This is only to demonstrate feasibility; it is not representative
of situations where there will be many observations away differing
from the training set (especially a synthetic one like \elasticc),
and one has to choose among hundreds of outliers / outlier clusters,
some of which are genuine rare / novel events, and some of which are
cases where the classifier was unsure / incorrect.

We tried two main approaches. The first consisted of fitting Gaussian
mixture models (GMMs) to embeddings, and looking at the likelihood
of held-out versus in-distribution examples. The intuition behind
this approach is that our training data should lie in a subspace of
the embedding, and novel objects should lie in another, and therefore
have relatively low probability in relation to the training data.
The results are shown in \Figref{GMM-anomaly-det}. Easier classes
like AGNs may be able to be picked out, but harder classes like CARTs
are not distinguishable.

Our second approach was to train an ATCAT variant to have a “possibly
novel class” output logit. (There are now 2 extra outputs: one
for “novel” and another for “unsure”.) For each batch
(256 LCs), we computed the mean and standard deviation of last layer
embeddings. These statistics were computed elementwise (so there are
384 of them), over batches. We then generated random activations matching
these means and standard deviations, and trained the final classifier
network to output “novel” for them. For the actual LCs, we trained
towards a KL loss with 91\% weight on the true label and 9\% on the
“unsure” logit. We then added another auxiliary loss towards
minimizing log-likelihood on the “novel” output class for actual
LCs, clamping at -10 log likelihood. These losses are intended to
make the model output “unsure” on points near its input domain,
and “novel” only on points farther from its input domain.

This approach was more successful, as shown in \Figref{Novel-det-model}.
AGNs, our easiest class, were detectable with near-100\% precision
at a good fraction of recall. For CARTs, our hardest-to-classify class
(see \Figref{Confusion-matrices-fine}), we only got near-100\% precision
at quite low precision. In a hypothetical situation where we wanted
to detect a new type of object like CARTs, this means that there would
have to be a lot of them, but it might be possible, especially if
we proceeded by studying the objects with highest “novel” score
first. However, this is a preliminary result. The fact that some CART
scores \emph{decreased} in “is novel” score when we increased
the aux loss suggests that this loss is not very robust. One possible
direction for future work is trying to use our model's generative
capabilities, generating improbable but not impossible light curves
via temperature sampling (see e.g. \citealt{Hinton_2015} for temperature
sampling).

\subsection{Discussion of model bias}\label{subsec:Discussion-bias}

We now briefly discuss model bias. The term “bias” has many
definitions; for example, in classical statistics, we might want to
estimate the mean of a distribution, and ensure that our estimator
on average (in expectation) returns the true mean. But this typically
doesn't make sense in classification, where we think of well-tuned
models hedging between classes when they are uncertain, and never
returning probabilities scores for a class that are negative or greater
than 1.

One type of bias we might be interested in is among the distribution
of classifier outputs themselves. For example, suppose we want to
estimate how many supernovae are Ib/c vs. Iax. If we force the classifier
to classify each object (even if it is unsure), then for our class-balanced
test distribution, our calibrated LC-only model will find that there
are $\approx1.7$x more Iax objects. Clearly, calibrating scores to
accuracy still results in class imbalances. Filtering to those with
$\geq80\%$ confidence score, our sample has $\approx1.67$x more
Iax objects. The more accurate LC+metadata model will find $\approx1.3$x
more Iax objects, and the imbalance gets larger with the $\geq80\%$
confidence score cutoff, now featuring $\approx1.6$x more Iax objects.
The effect is loosely correlated with classifier score: classes with
very good precision/recall will have counts near their true class
counts, but other classes may be above and below, and the distance
isn't strictly correlated with F1 score.

Another type of bias is the conferred bias when making quality cuts
using our model. It is very common to study a property of objects
by making quality cuts, see \citet{Babusiaux2018_ObservationalHRD}
for example. Let's call this property $\mathfrak{A}$; it could be
numeric (e.g. estimated mass) or binary (e.g. is the mass less than
some threshold). The key question is: if we select objects using the
classifier outputs (say everything matching class $c$ with confidence
$s$), how will $\mathfrak{A}$ change? We propose that the risk depends
on a few factors: the base accuracy of the classifier, how $\mathfrak{A}$
appears in mis-classified objects, and $\mathfrak{A}$'s correlation
with classification scores. At the limit where the classifier is very
good, then our quality cut will be equal to the true sample, and the
other concerns are not relevant. Otherwise, we will be choosing a
model score threshold $s$ with a certain precision-recall tradeoff
(often visualized with a precision-recall curve; we recommend Chapter
8 of \citealp{manning2008introduction} to unfamiliar readers). Higher
recall selections (low value of $s$) will select many objects of
different classes, so the result of our study will mostly depend on
how $\mathfrak{A}$ manifests in mis-classified objects. If $\mathfrak{A}$
is extremely different in mis-classified objects, then these outliers
can be filtered out; conversely, if $\mathfrak{A}$ has the same distribution
in mis-classified objects, then the result of our study won't change.
So the most adversarial case lies in the middle. Higher precision
selections (high value of $s$), by contrast, will depend more on
the correlation of $\mathfrak{A}$ with the classification score.
Particularly, if some settings of $\mathfrak{A}$ make it easier or
harder to classify, then our high precision sample will be skewed.

In Appendix \ref{subsec:Empirical-bias-study}, we studied this empirically
on two simulator parameters, \texttt{LOGMASS\_TRUE} and \texttt{LOG\_SFR}
(removed from our datasets by default). We tried to find more interesting/varying
parameters, but were unable to find ones that had terribly pathological
behavior, in part because these parameters are only present on some
classes of objects. We studied the property over different precision/recall
tradeoffs, since high-precision and high-recall selections may be
affected through different phenomena.

Causes and ameliorations of bias also vary. One serious issue is domain
adaptation. When a model is trained on one domain and evaluated on
another, then this can not only harm its overall accuracy, but lead
to objects of one type systemically being classified as another. \citet{Richards_2012}
studies this effect on variable star classification on the OGLE (Optical
Gravitational Lensing Experiment) dataset, and finds that active learning
is quite effective.

Future strategies for amelioration may also draw inspiration from
the large body of ML fairness work for analysis and potential solutions
to the problem (see for example \citealp{D'Amour_2020,Ayres_2002_outcome_tests_fairness,Davies_2017_fairness}).
Perhaps surprisingly, the statistical structure of the latter type
of bias we discuss appears to be similar to certain fairness problems.
ML fairness deals with problems like correlation of skin tone (analogous
to our “property $\mathfrak{A}$”, often not labeled in the
original training data) and skin condition classifier performance
(analogous to our classifier output).

To conclude, we strongly encourage practitioners to find a reasonable
statistical setup for their problem, instead of simply using our classifier
outputs as priors, even after calibration. We should be specific about
what calibration tries to achieve,
\[
P_{X,Y\sim\mathcal{D}}\left(\left.Y=\underset{c}{\text{argmax}}f_{c}\left(X\right)\right|\max_{c}f_{c}\left(X\right)\approx s\right)\approx s
\]
e.g. if the classifier confidence (top output score) is around 90\%
($s=0.9$), then for a well-calibrated model on a representative dataset,
90\% of these examples will have the correct label. Calibration does
not provide a guarantee about the balance of top classifications,
or how likely the classifier is to be confident of examples in a particular
class. It is, however, useful for estimating the precision of a sample,
for situations when we are trying to manage the model's precision-recall
tradeoff.

\subsection{Future work: Generative possibilities}\label{subsec:FW-generative}

Since our model has generative capabilities (it predicts the next
flux given the time, channel wavelength, and flux\_err) and can classify
such LCs, it would be valuable to look at expected informational gain.
One way of defining this could be
\[
\text{E}_{x_{t+1}}\left[\text{KL}\left(q\left(Y|x_{t+1},x_{t},...\right)||q\left(Y|x_{t},...\right)\right)\right]
\]
following \citet{Li_2024}. This could be used to evaluated proposed
sequences of observations; for example, a telescope path and color
filter selection might specify time and channel wavelength values,
and then flux\_err values could drawn from a simple distribution (e.g.
sampling randomly from previous flux errors). Then, outer expectation
could be approximated by sampling next flux values from the model's
generative output. The inner KL term is the discrete multinomial KL
divergence; before adding the new point, the classifier gives us a
vector of probabilities over classes ($q\left(Y|x_{t},...\right)$),
and we compare that to the probabilities over classes after the new
point has been added ($q\left(Y|x_{t+1},x_{t},...\right)$). However,
there are probably many cases where one has technically gained information,
but the distinctions aren't as interesting, for example one experimenter
might not care about distinguishing RR-Lyrae vs. Delta Scuti's, but
might care a lot about CARTs vs. Type-II supernovae. In that case,
collapsing classes, or choosing another metric than KL altogether
would be most valuable.

\subsection{Future work: Further modeling improvements}\label{subsec:FW-modeling}

Combining ATCAT's architecture with other techniques should also be
a fruitful avenue of exploration.

Coarse-grained classifications may be useful, especially for early
detection. These can both help to focus on classes / class boundaries
of more specific interest (for example, focusing on rare transients
such as KN, TDE, etc. and excluding more common ones such as Type
Ia and variable stars), and also to achieve higher accuracy via hierarchical
losses, as demonstrated by ORACLE \citep{Shah_2025}.

Applying ATCAT to other surveys, and training a cross-survey variant
would also be interesting and potentially valuable. We would specifically
like to combine ZTF and LSST data. We hope our work making the model
work on only channel wavelengths instead of integer channel indices
(which required a change to the embedder) will be helpful. A cross-survey
model could also be combined with expected informational gain metrics
(see Section \ref{subsec:FW-generative}) to prioritize follow-up
observations with different telescopes. We would again like to stress
the value of better dataset formats and standardization (Section \ref{subsec:FW-Better-standardization});
in this case it is also critical to have light curves assembled reflecting
the same source observed in multiple channel wavelengths, since it
is a prohibitive amount of work to simultaneously do this and build
cross-survey models.

Finally, low-level improvements in the ATCAT architecture may be valuable.
Despite our advances in modeling accuracy, we believe we have not
removed all of the headroom. For LC encoding, \citet{Su_2021} also
found that applying the rotary encoder in the attention mechanism
was better. We would also like to explore the effects of repeated
observations; while it is not a-priori bad that our model is sensitive
to them (they may reflect increased confidence, for example), its
sensitivity may differ from convolutional approaches. In Section \ref{subsec:model-arch-ablation}
we saw slightly lower LC-only performance on larger models, which
indicates sub-optimal regularization, initialization, or architecture
choices. For early detection, we found that better settings for early
detection resulted in a slight loss of accuracy on classifications
after 1024 days since the initial alert. A “composite” model
which switches between our base models and early detection models
(based on the number of points) would be trivially better, indicating
some modeling, loss, or training headroom. We also believe that more
advanced schemes for weighting the loss could be useful, if it could
be deduced at which LC points the model should be able to make a classification,
and which ones are just adding noise. The early detection loss model
is also a bit slower to train, possibly due to larger all-reduce steps
in the gradient calculation.

\subsection{Future work: Further performance improvements}\label{subsec:FW-throughput-perf}

Despite our competitive showing in training and inference performance,
we believe significant headroom still exists. We suggest investigating
further sequence length reductions primarily. This could come as simply
truncating / subsampling long sequences, or doing a more sophisticated
sequence packing scheme. Our current masking scheme is quite wasteful.
The length of \elasticc sequences are quite imbalanced; with our
/ ATAT's splits, in the training set, only \res{train percent non-masked sequence elements}\%
of sequence elements are \emph{not} masked on average, and in the
test set, which has even sampling of classes, \res{test percent non-masked sequence elements}\%
were not masked. So this is a huge potential speedup for both training
and inference. Unfortunately, sequence packing does usually have a
medium-high code complexity cost to logic in datasets, models, and
loss functions. We also suggest experimenting with a hard restriction
on number of LC points (say, $k$) local attention layers can attend
to, changing the attention time complexity from $\mathcal{O}\left(n^{2}\right)$
to $\mathcal{O}\left(kn\right)$ for those layers. We use the \texttt{flex\_attention}
library \citep{Dong_2024}, which claims to optimize for this case
(called “sliding window attention”).

Furthermore, if inference performance of ensembled models is important,
then it should be investigated whether the results can be distilled
into a single model (e.g. using KL divergence towards silver labels).
We also did preliminary experiments with stochastic weights averaging,
but got inferior results to ensembling; further investigation may
be reasonable.

ATCAT is a small enough model that it should be possible to make it
run reasonably on the CPU. Additionally, several CPUs, including popular
laptop CPUs, now include a small integrated GPU or NPU (neural processing
unit), which could be leveraged using PyTorch backends MPS for Apple's
M-series chips, ROCm for AMD's APUs, or OpenCL for several platforms.

\subsection{Future work: Better dataset standardization}\label{subsec:FW-Better-standardization}

As mentioned in Section \Subsecref{Main-results}, it is unfortunate
that there are many variants of \elasticc preprocessing, which makes
it difficult to compare models. Both for \elasticc and other surveys,
several strategies may help facilitate these comparisons. We suggest
the following,
\begin{itemize}
\item Using Parquet files,\footnote{Any Parquet-compatible implementation can be used, although as of
writing Pandas notably lacks support for lists, and we suggest it
not be prioritized given Polars' vastly superior performance and more
principled API.} where each row is a LC. In the course of this work, we experimented
with Dask, Lance, and Polars, and \textbf{highly recommend use of
Polars and Parquet}. We suggest sharding and zstandard compression;
with the right shard size, this provides good compression and multi-threaded
loading. Metadata become “normal” columns, and LC points can
become list-type columns.
\item Ensuring all fields are well-documented, including their units, and
methodologies and rationale behind upstream processing (such as mean
field subtraction). If it is expected that many practitioners will
filter by a particular bitmask\footnote{Parquet supports boolean (single bit) or enum values, which may be
cleaner than traditional integer bitmasks. Flags can also be combined
into a struct, for grouping or reducing the number of top-level columns.} / flag, then clean, idiomatic Python code for that should be provided.
\item Making datasets usable with minimal preprocessing for basic tasks.
When there are good defaults (such as filtering by flags), we suggest
applying those to the main dataset, and separately sharing dataset
variants without these defaults. If issues arise, such as SNIDs being
duplicated between classes, then there should be a mechanism for fixing
this at the source, perhaps using minor version numbers (e.g. “v3.1”).
\item Declaring a standard schema for expressing training / validation /
test splits, and collapsing of fine classes (labels). We suggest Parquet
tables with SNIDs for splits (a balance of being explicit and efficient),
and either a Parquet table, or clean, idiomatic Python code expressing
fine classes collapsing logic. Different astronomers will care about
modeling different aspects, so collapsing classes is natural and shouldn't
be dictated by the original dataset providers, but schemas / standardization
would make it easier to directly compare models.
\item Having clear licensing terms, specifying whether / how derivative
datasets can be distributed.
\end{itemize}
We are hesitant to provide these solutions ourselves, to avoid proliferating
standards.\footnote{\url{https://xkcd.com/927/}} We hope that the
above suggestions can be received as constructive criticism for anyone
releasing datasets. Our criticism is not directed at \elasticc as
much as the state of \elasticc classification \textasciitilde 2
years after its release. Clearly, our work would not have been possible
without the work of the \elasticc team, who assembled a dataset from
a large variety of simulators; their work has realized immense scientific
value for the LSST, and will continue to do so.

\section{Acknowledgments}

I would like to thank the ATAT team for making their code and results
publicly available, including all necessary components to reproduce
the results in their paper. I would also like to thank Josh Bloom
for many illuminating discussions, and inviting me to join discussions
with his research group.

\bibliographystyle{mnras}
\bibliography{citations}

\begin{appendices}

\section{Appendix}

\subsection{Class exemplars and “mis-exemplars”}

\begin{figure*}
\begin{centering}
\includegraphics[width=2\columnwidth]{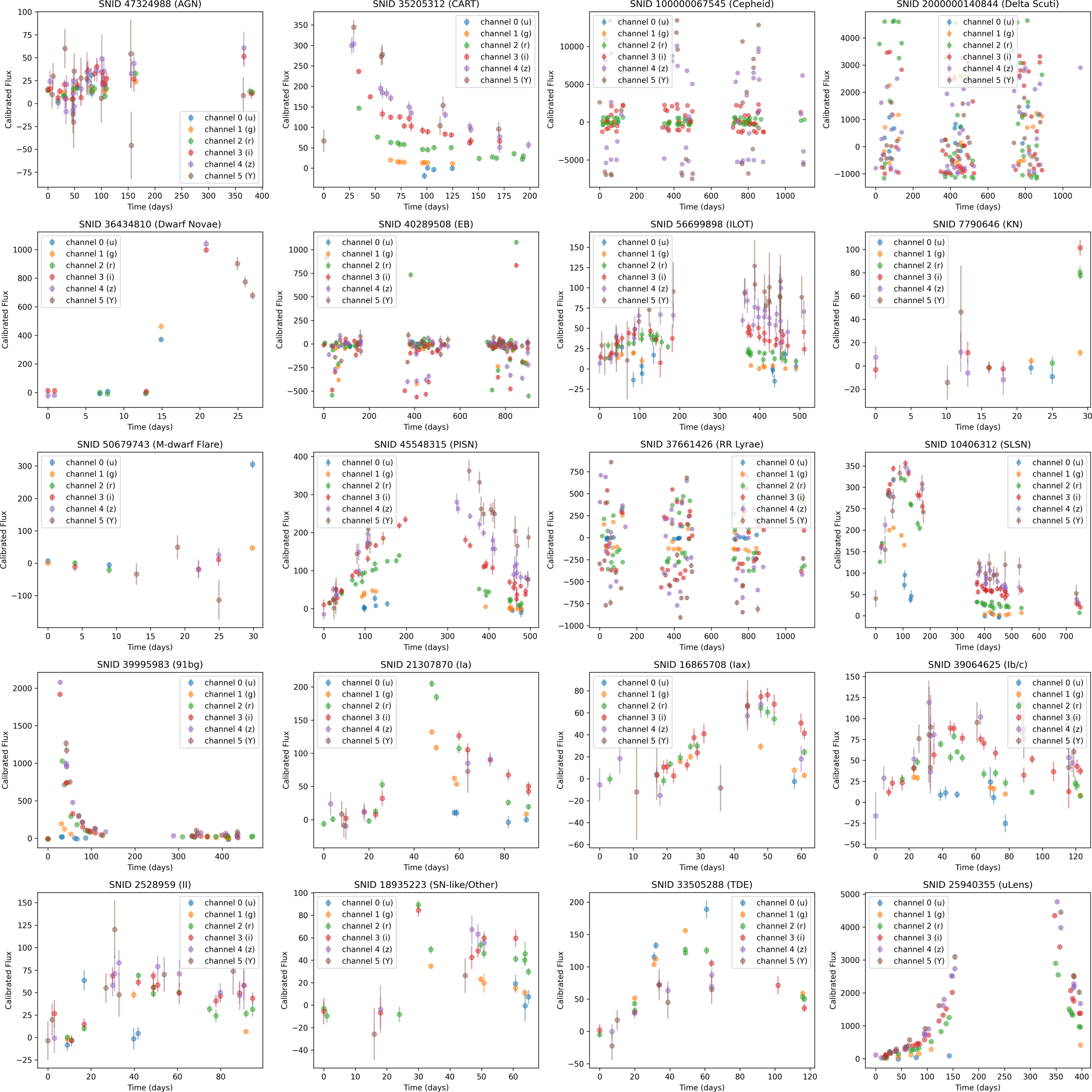}
\par\end{centering}
\caption{\textbf{Class exemplars}. For each class (using the ATAT 20-way classification
scheme), we found the highest-scoring example in our validation set,
using our LC-only model.}\label{fig:Class=000020exemplars}
\end{figure*}

\begin{figure*}
\begin{centering}
\includegraphics[width=2\columnwidth]{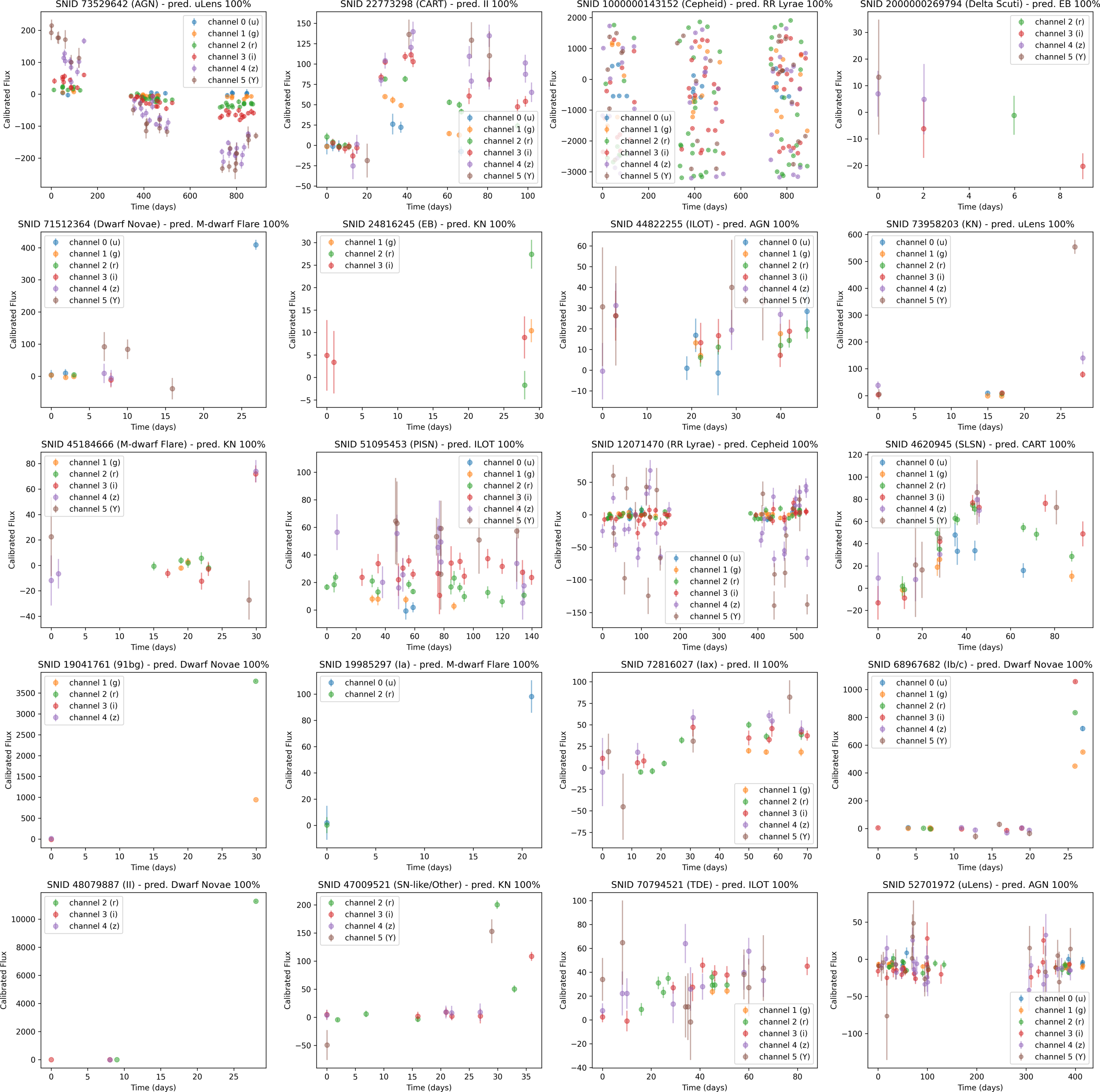}
\par\end{centering}
\caption{\textbf{Examples strongly misclassified by our model} (“class
mis-exemplars”). For each class (using the ATAT 20-way classification
scheme), we found the example in our validation set that scored the
highest on \emph{another} label, using our LC-only model. We can see
that there is some trivial headroom in terms of the variable stars
(RR-Lyrae, Delta-Scuti, Cepheid), and likely making our model incorporate
period folding and/or Blazhko RR-Lyrae detection would improve those
substantially. However, their F-1 scores are generally quite high.
We did some preliminary experiments period folding all light curves
and did not see a benefit, but believe this is because we need to
do significance testing before period folding (and/or detect a distribution
of likely periods).}\label{fig:Class-misexemplars}
\end{figure*}
\begin{figure*}
\begin{centering}
\includegraphics[width=2\columnwidth]{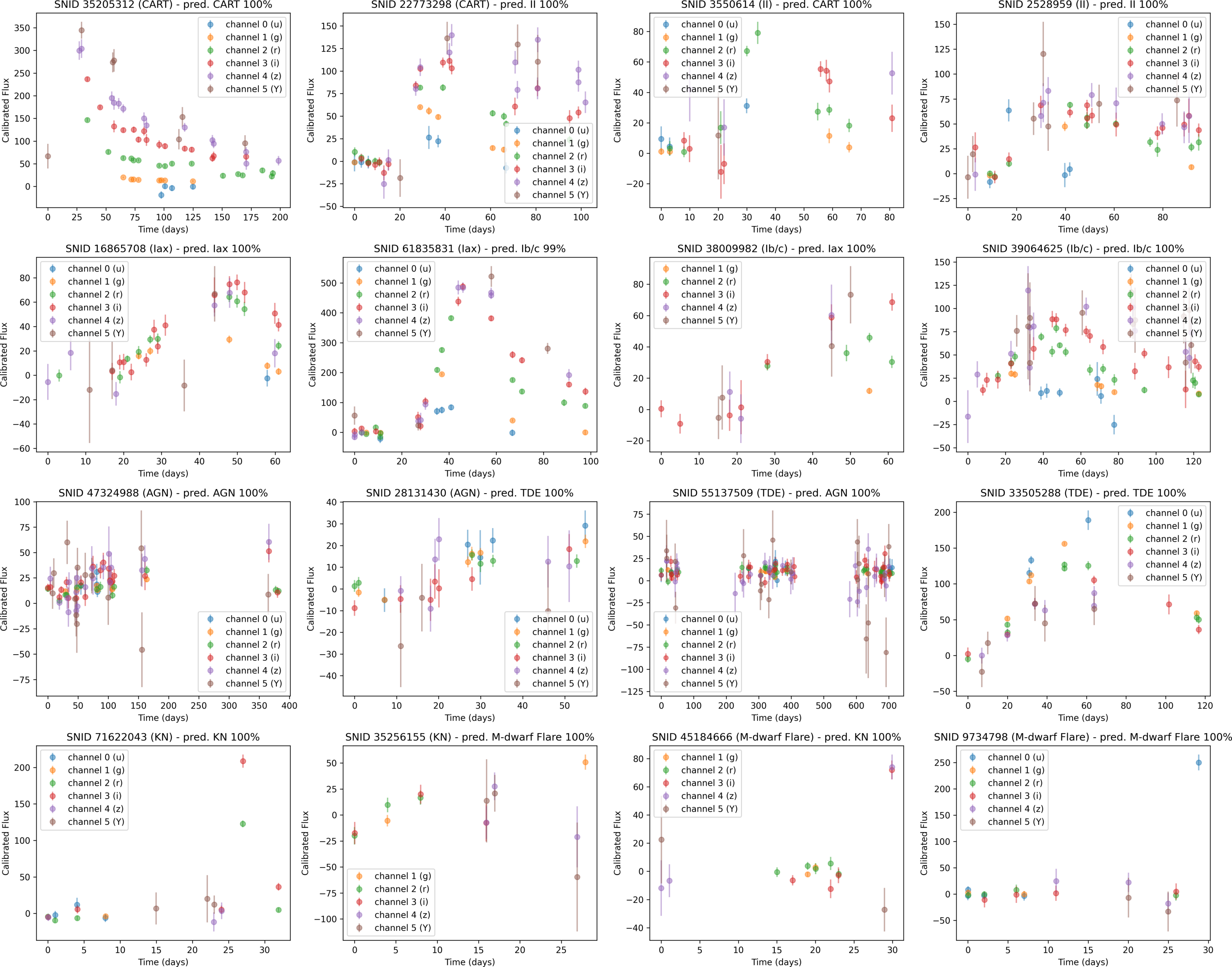}
\par\end{centering}
\caption{\textbf{Pairwise confusions between classes}. On the left column,
we have the best class exemplar for one class $i$, and on the right
column, we have the best class exemplar for another class $j$. Then,
in the second (middle left) column, we have an exemplar from true
class $i$ that scored highest for class $j$, and in the third (middle
right) column, we have an exemplar from true class $j$ that scored
the highest for class $i$.}\label{fig:Pairwise-confusions}
\end{figure*}

We provide several example light curves, to help make our classification
problem concrete. In \Figref{Class=000020exemplars} we show class
exemplars, or examples in each class where our LC-only model gave
the example a high score. In \Figref{Class-misexemplars} we show
“mis-exemplars”, examples where our model gave a very low score
to the true class. And in \Figref{Pairwise-confusions} we specifically
examine some pairs of classes which had high confusion scores in the
confusion matrices (\Figref{Confusion-matrices-fine}), visualizing
exemplars from each class on the left and right, and then confusions
in the middle two columns (examples from the left class classified
as the right class, and examples from the right class classified as
the left).

\subsection{Reasonable scaling of ATAT's single-core CPU results}\label{subsec:multicore-CPU-scale}

ATAT  reported their feature extraction CPU performance as a single-core
number, but since processing batches of light curves is trivially
parallelizable, we did a back-of-the-envelope calculation to scale
it to a reasonable multi-core number.

\emph{Naive calculation and why it's wrong}: One might be tempted
to multiply the performance by the number of cores in this chip (64),
and then upscale by the nominal power consumption of the GPU ($\nicefrac{400}{225}$),
achieving a scalar of 114x. However, CPUs almost never linearly scale
from single to multi-core performance, because (a) especially relevant
for scientific tasks, the number of AVX and SSE units is often limited,
resulting in reduced instruction-level parallelism; the chips also
often downclock  when many of these are used in parallel (b) L1,
L2, and L3 caches are limited and experience more pressure in multi-core
workloads. The EPYC 7662 is also a newer chip than the A100.

\emph{Our calculation}: The Ryzen 3995WX is a well-regarded CPU of
the same generation as the A100, so we took its multi-core score according
to cpubenchmark.net (83956), and divided it by the single-core score
of the EPYC 7662 (2102), and then scaled by the nominal power consumption
of the GPU ($\nicefrac{400}{225}$), resulting in 57x scaling. We
should also note that while we can monitor the GPU's power consumption,
we don't know what the CPU's would be like in this theoretical workload.
CPUs report a “Typical TDP” which is sometimes less than their
max/peak TDP in heavy workloads.

\subsection{Empirical evaluation of model bias on simulator parameters}\label{subsec:Empirical-bias-study}

\begin{figure*}
\begin{centering}
\includegraphics[totalheight=9cm]{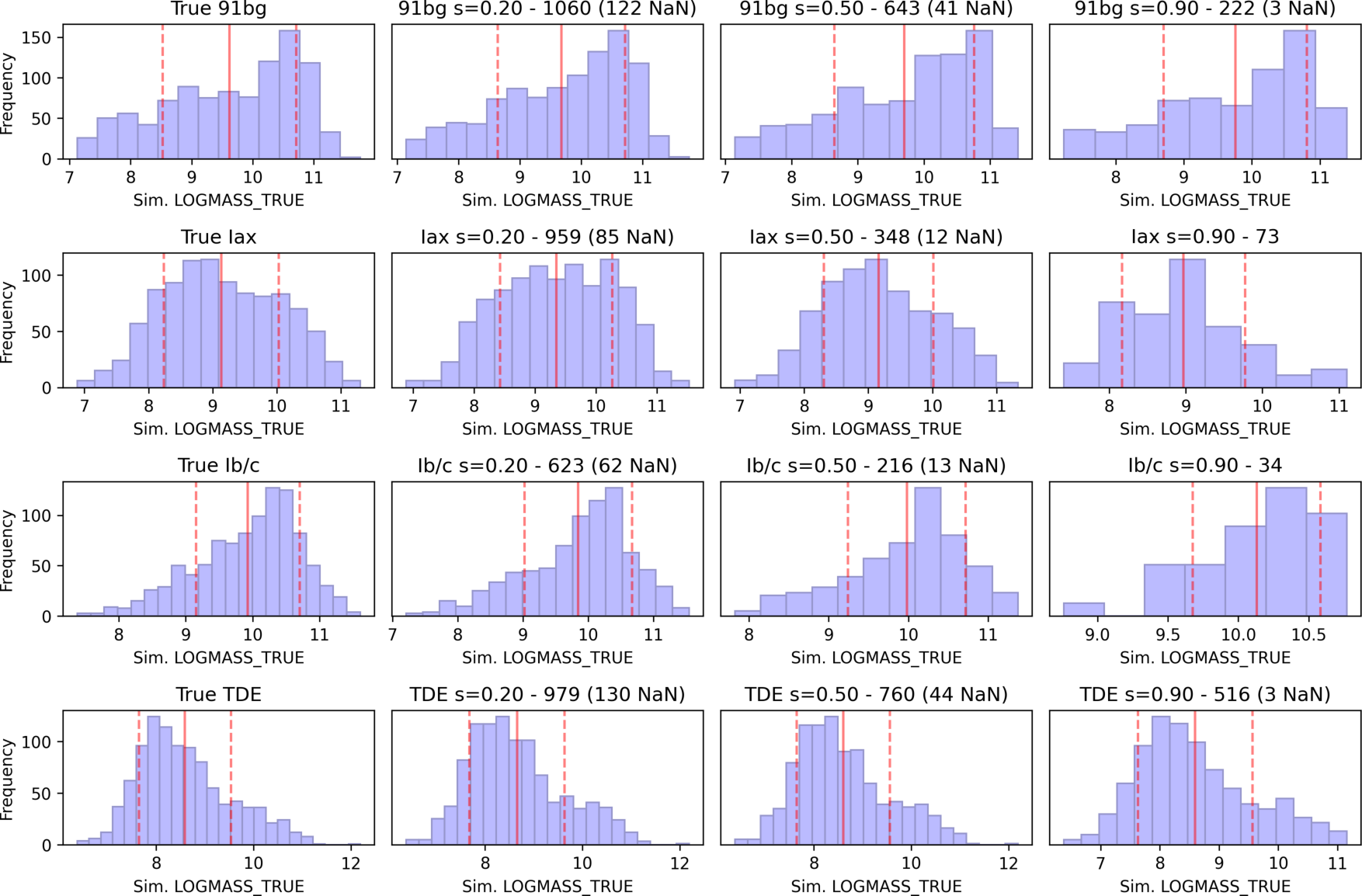}~~\includegraphics[totalheight=9cm]{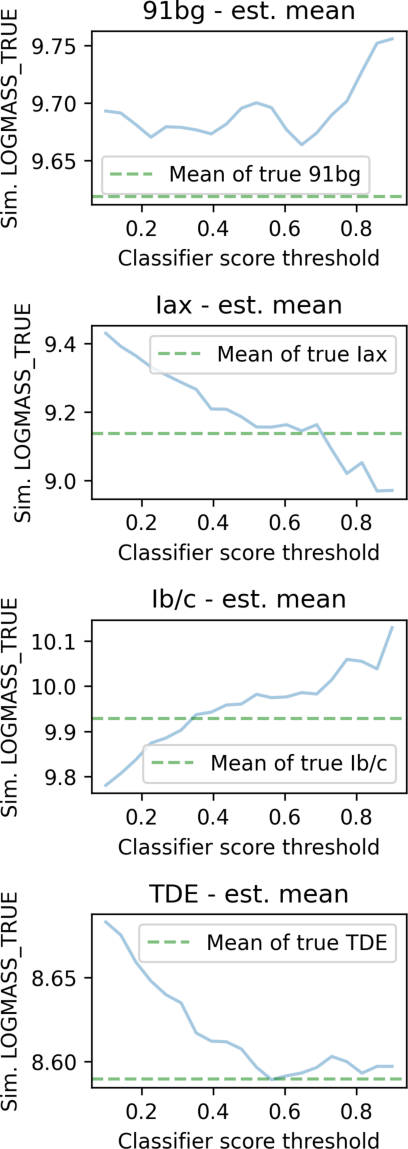}
\par\end{centering}
\caption{\textbf{Empirical study of model selection bias effects on simulator
parameter LOGMASS\_TRUE}. Left: True label distributions. Middle 3
columns: distributions when selecting by classifier probability score,
using our calibrated LC-only model. Right: mean estimate vs. score
threshold. The “s=” is the minimum score threshold, and after
the “-” the total number of objects selected is shown. In parenthesis
is the number of NaNs (from classes without LOGMASS\_TRUE values),
which we removed for this preliminary experiment.}\label{fig:Empirical-logmass-bias}
\end{figure*}

\begin{figure*}
\begin{centering}
\includegraphics[totalheight=9cm]{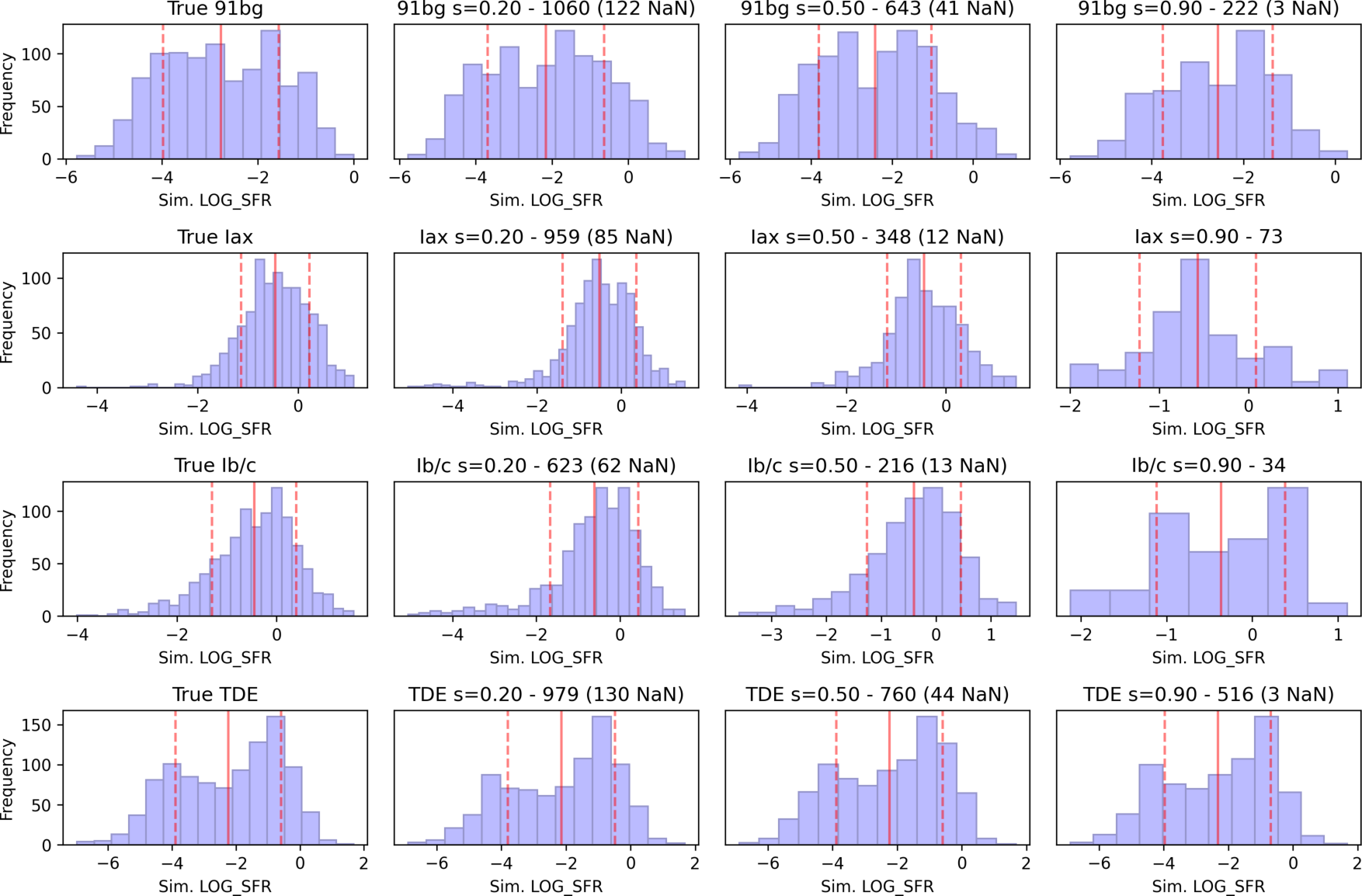}~~\includegraphics[totalheight=9cm]{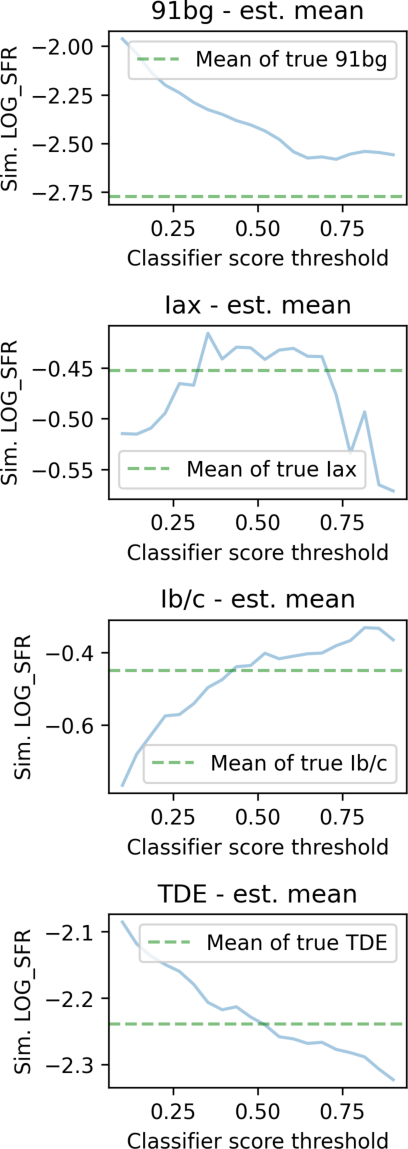}
\par\end{centering}
\caption{\textbf{Empirical study of model selection bias effects on simulator
parameter LOG\_SFR}. Please see \Figref{Empirical-logmass-bias} for
the format of this figure.}\label{fig:Empirical-sfr-bias}
\end{figure*}

In \Figref{Empirical-logmass-bias} and \Figref{Empirical-sfr-bias},
we examine the bias in two \elasticc simulator parameters, \texttt{LOGMASS\_TRUE}
and \texttt{LOG\_SFR}, imparted by using quality cuts on our calibrated
LC-only model, instead of the true label. These parameters appeared
to be shared by several supernova simulators, although they did not
appear for many other classes. For this preliminary investigation,
we simply removed the NaN values (corresponding to classes without
these parameters). We would argue that the most adversarial values
for the parameter on incorrectly-classified examples must not be extreme
outliers (or else we could easily remove them), but still bias the
distribution. Generally, these two parameters don't suffer from much
bias, despite having different distributions on classes and the fact
that we used the LC-only model. The distribution shapes do not appear
to change significantly, and the estimated mean from different samples
is generally fairly controlled. The 91bg class showed sample means
which were a small bit off from the true mean on both of these parameters.
It would be valuable future work to see if different random initializations
(or bootstrappings) of this model would have the same consistent pattern.
Please see Section \ref{subsec:Discussion-bias} for general bias
discussion.

\end{appendices}
\end{document}

%% file: results/macros.tex
\ExplSyntaxOn
\prop_new:N \g_results_prop

\NewDocumentCommand{\defineresult}{mm}{
  \prop_gput:Nnn \g_results_prop {#1} {#2}
}

\NewDocumentCommand{\res}{m}{
  \prop_item:Nn \g_results_prop {#1}
}
\ExplSyntaxOff

\defineresult{atat lc_only f1}{62.7 $\pm$ 0.4}
\defineresult{atat lc_meta f1}{83.5 $\pm$ 0.6}
\defineresult{our lc_only f1}{71.79 $\pm$ 0.28}
\defineresult{our lc_meta f1}{89.75 $\pm$ 0.04}
\defineresult{abl data aug lc_only baseline}{69.1}
\defineresult{abl data aug lc_only only random noise}{69.7}
\defineresult{abl data aug lc_only only flux scale}{69.2}
\defineresult{abl data aug lc_only only redshift}{70.1}
\defineresult{abl data aug lc_only only subsample}{70.7}
\defineresult{abl data aug lc_only only time shift}{70.4}
\defineresult{abl data aug lc_only scheme1}{72.1}
\defineresult{abl data aug lc_only scheme3}{71.4}
\defineresult{abl data aug lc_meta baseline}{88.8}
\defineresult{abl data aug lc_meta only random noise}{88.5}
\defineresult{abl data aug lc_meta only flux scale}{88.8}
\defineresult{abl data aug lc_meta only redshift}{88.8}
\defineresult{abl data aug lc_meta only subsample}{89.3}
\defineresult{abl data aug lc_meta only time shift}{89.1}
\defineresult{abl data aug lc_meta scheme1}{89.5}
\defineresult{abl data aug lc_meta scheme3}{89.6}
\defineresult{abl emb lc_meta current_rotary}{89.86 $\pm$ 0.1}
\defineresult{abl emb lc_meta legacy_atat}{79.79 $\pm$ 0.42}
\defineresult{abl emb lc_meta legacy_atat_with_tanh}{83.3 $\pm$ 0.04}
\defineresult{abl emb lc_meta pre_rotary}{89.5 $\pm$ 0.07}
\defineresult{abl emb lc_meta pre_rotary_no_scaling}{89.37 $\pm$ 0.13}
\defineresult{abl emb lc_meta pre_rotary_no_scaling_no_ch_no_flux_err}{84.7 $\pm$ 0.12}
\defineresult{abl emb lc_meta pre_rotary_no_scaling_no_flux_err}{88.86 $\pm$ 0.09}
\defineresult{abl emb lc_only current_rotary}{71.87 $\pm$ 0.14}
\defineresult{abl emb lc_only legacy_atat}{61.53 $\pm$ 0.76}
\defineresult{abl emb lc_only legacy_atat_with_tanh}{57.94 $\pm$ 0.13}
\defineresult{abl emb lc_only pre_rotary}{71.73 $\pm$ 0.05}
\defineresult{abl emb lc_only pre_rotary_no_scaling}{71.13 $\pm$ 0.24}
\defineresult{abl emb lc_only pre_rotary_no_scaling_no_ch_no_flux_err}{60.85 $\pm$ 0.12}
\defineresult{abl emb lc_only pre_rotary_no_scaling_no_flux_err}{69.95 $\pm$ 0.07}
\defineresult{abl model lc_meta baseline}{89.66 $\pm$ 0.14}
\defineresult{abl model lc_meta larger}{89.85 $\pm$ 0.05}
\defineresult{abl model lc_meta non_hybrid}{89.29 $\pm$ 0.17}
\defineresult{abl model lc_only baseline}{71.87 $\pm$ 0.07}
\defineresult{abl model lc_only larger}{71.69 $\pm$ 0.19}
\defineresult{abl model lc_only non_hybrid}{71.86 $\pm$ 0.12}
\defineresult{abl loss opt lc_meta baseline}{89.91 $\pm$ 0.1}
\defineresult{abl loss opt lc_meta default_adam_no_sched}{89.02 $\pm$ 0.1}
\defineresult{abl loss opt lc_meta f1_only}{87.77 $\pm$ 0.16}
\defineresult{abl loss opt lc_meta logloss_only}{89.57 $\pm$ 0.04}
\defineresult{abl loss opt lc_meta no_extra_output_class}{89.79 $\pm$ 0.09}
\defineresult{abl loss opt lc_only baseline}{71.79 $\pm$ 0.15}
\defineresult{abl loss opt lc_only default_adam_no_sched}{71.47 $\pm$ 0.11}
\defineresult{abl loss opt lc_only f1_only}{70.53 $\pm$ 0.11}
\defineresult{abl loss opt lc_only logloss_only}{71.51 $\pm$ 0.14}
\defineresult{abl loss opt lc_only no_extra_output_class}{71.81 $\pm$ 0.04}
\defineresult{perfbench atat feat extraction}{5.3}
\defineresult{perfbench atat feat extraction scaled}{303}
\defineresult{perfbench atat model no feats}{2105}
\defineresult{perfbench atat model with feats}{1206}
\defineresult{perfbench oracle}{7692}
\defineresult{perfbench our model lc only 40gb}{41905}
\defineresult{perfbench our model lc only 80gb}{45568}
\defineresult{perfbench our model lc meta 40gb}{40283}
\defineresult{perfbench our model lc meta 80gb}{44273}
\defineresult{perfbench our model lc meta h100}{68906}
\defineresult{perfbench our model lc meta rtx 4090}{33153}
\defineresult{perfbench our model lc meta rtx 4090 rounded}{33000}
\defineresult{our model times faster vs atat}{174}
\defineresult{our model times faster vs atat gpu model only}{35}
\defineresult{our model times faster vs oracle}{9}
\defineresult{our model times faster on h100}{1.6}
\defineresult{pretr true baseline lc_only f1}{63.1 $\pm$ 0.6}
\defineresult{pretr true baseline lc_meta f1}{83.7 $\pm$ 0.1}
\defineresult{pretr no ens lc_only f1}{65.5 $\pm$ 0.1}
\defineresult{pretr no ens lc_meta f1}{85.7 $\pm$ 0.4}
\defineresult{pretr ens lc_only f1}{66.73 $\pm$ 0.15}
\defineresult{pretr ens lc_meta f1}{86.66 $\pm$ 0.13}
\defineresult{pretr setup A lc_only f1}{67.44 $\pm$ 0.06}
\defineresult{pretr setup A lc_meta f1}{87.07 $\pm$ 0.18}
\defineresult{our lc_meta num params}{4.1M}
\defineresult{our train size}{1.5M}
\defineresult{our val size}{365K}
\defineresult{our test size}{20K}
\defineresult{train percent non-masked sequence elements}{17.7}
\defineresult{test percent non-masked sequence elements}{26.0}

%% file: paper.bbl
\begin{thebibliography}{}
\makeatletter
\relax
\def\mn@urlcharsother{\let\do\@makeother \do\$\do\&\do\#\do\^\do\_\do\%\do\~}
\def\mn@doi{\begingroup\mn@urlcharsother \@ifnextchar [ {\mn@doi@}
  {\mn@doi@[]}}
\def\mn@doi@[#1]#2{\def\@tempa{#1}\ifx\@tempa\@empty \href
  {http://dx.doi.org/#2} {doi:#2}\else \href {http://dx.doi.org/#2} {#1}\fi
  \endgroup}
\def\mn@eprint#1#2{\mn@eprint@#1:#2::\@nil}
\def\mn@eprint@arXiv#1{\href {http://arxiv.org/abs/#1} {{\tt arXiv:#1}}}
\def\mn@eprint@dblp#1{\href {http://dblp.uni-trier.de/rec/bibtex/#1.xml}
  {dblp:#1}}
\def\mn@eprint@#1:#2:#3:#4\@nil{\def\@tempa {#1}\def\@tempb {#2}\def\@tempc
  {#3}\ifx \@tempc \@empty \let \@tempc \@tempb \let \@tempb \@tempa \fi \ifx
  \@tempb \@empty \def\@tempb {arXiv}\fi \@ifundefined
  {mn@eprint@\@tempb}{\@tempb:\@tempc}{\expandafter \expandafter \csname
  mn@eprint@\@tempb\endcsname \expandafter{\@tempc}}}

\bibitem[\protect\citeauthoryear{Ansari et~al.,}{Ansari
  et~al.}{2024}]{Ansari_2024}
Ansari A.~F.,  et~al., 2024, arXiv preprint arXiv:2403.07815v3

\bibitem[\protect\citeauthoryear{Ayres}{Ayres}{2002}]{Ayres_2002_outcome_tests_fairness}
Ayres I.,  2002, \mn@doi [Justice Research and Policy]
  {10.3818/JRP.4.1.2002.131}, 4, 131

\bibitem[\protect\citeauthoryear{Bailey, Aragon, Romano, Thomas, Weaver  \&
  Wong}{Bailey et~al.}{2007}]{Bailey_2007}
Bailey S.,  Aragon C.,  Romano R.,  Thomas R.~C.,  Weaver B.~A.,   Wong D.,
  2007, arXiv preprint arXiv:0705.0493

\bibitem[\protect\citeauthoryear{Bender, Gebru, McMillan-Major  \&
  Shmitchell}{Bender et~al.}{2021}]{Bender_Gebru_Stochastic_Parrots}
Bender E.~M.,  Gebru T.,  McMillan-Major A.,   Shmitchell S.,  2021, in
  Proceedings of the 2021 ACM Conference on Fairness, Accountability, and
  Transparency. FAccT '21.
Association for Computing Machinery, New York, NY, USA, p. 610–623,
  \mn@doi{10.1145/3442188.3445922}, \url
  {https://doi.org/10.1145/3442188.3445922}

\bibitem[\protect\citeauthoryear{{Bloom} et~al.,}{{Bloom}
  et~al.}{2012}]{Bloom_2012_aut_disc}
{Bloom} J.~S.,  et~al., 2012, \mn@doi [Publications of the Astronomical Society
  of the Pacific] {10.1086/668468}, \href
  {https://ui.adsabs.harvard.edu/abs/2012PASP..124.1175B} {124, 1175}

\bibitem[\protect\citeauthoryear{Boone}{Boone}{2019}]{Boone_2019}
Boone K.,  2019, arXiv preprint arXiv:1907.04690v2

\bibitem[\protect\citeauthoryear{Boone}{Boone}{2021}]{Boone_2021}
Boone K.,  2021, \mn@doi [The Astronomical Journal] {10.3847/1538-3881/ac2a2d},
  162, 275

\bibitem[\protect\citeauthoryear{Cabrera-Vives et~al.,}{Cabrera-Vives
  et~al.}{2024}]{Cabrera-Vives_2024}
Cabrera-Vives G.,  et~al., 2024, arXiv preprint arXiv:2405.03078v2

\bibitem[\protect\citeauthoryear{{Carrasco-Davis} et~al.,}{{Carrasco-Davis}
  et~al.}{2019}]{Carrasco-Davis_2019}
{Carrasco-Davis} R.,  et~al., 2019, \mn@doi [Publications of the Astronomical
  Society of the Pacific] {10.1088/1538-3873/aaef12}, \href
  {https://ui.adsabs.harvard.edu/abs/2019PASP..131j8006C} {131, 108006}

\bibitem[\protect\citeauthoryear{{Charnock} \& {Moss}}{{Charnock} \&
  {Moss}}{2017}]{Charnock_2017}
{Charnock} T.,  {Moss} A.,  2017, \mn@doi [Astrophysical Journal Letters]
  {10.3847/2041-8213/aa603d}, \href
  {https://ui.adsabs.harvard.edu/abs/2017ApJ...837L..28C} {837, L28}

\bibitem[\protect\citeauthoryear{Corbett-Davies, Pierson, Feller, Goel  \&
  Huq}{Corbett-Davies et~al.}{2017}]{Davies_2017_fairness}
Corbett-Davies S.,  Pierson E.,  Feller A.,  Goel S.,   Huq A.,  2017, in
  Proceedings of the 23rd ACM SIGKDD International Conference on Knowledge
  Discovery and Data Mining. KDD '17.
Association for Computing Machinery, New York, NY, USA, p. 797–806,
  \mn@doi{10.1145/3097983.3098095}, \url
  {https://doi.org/10.1145/3097983.3098095}

\bibitem[\protect\citeauthoryear{D'Amour et~al.,}{D'Amour
  et~al.}{2020}]{D'Amour_2020}
D'Amour A.,  et~al., 2020, arXiv preprint arXiv:2011.03395v2

\bibitem[\protect\citeauthoryear{Dong, Feng, Guessous, Liang  \& He}{Dong
  et~al.}{2024}]{Dong_2024}
Dong J.,  Feng B.,  Guessous D.,  Liang Y.,   He H.,  2024, arXiv preprint
  arXiv:2412.05496

\bibitem[\protect\citeauthoryear{{Donoso-Oliva}, {Becker}, {Protopapas},
  {Cabrera-Vives}, {Vishnu}  \& {Vardhan}}{{Donoso-Oliva}
  et~al.}{2023}]{Donoso-Oliva_2003_astromerv1}
{Donoso-Oliva} C.,  {Becker} I.,  {Protopapas} P.,  {Cabrera-Vives} G.,
  {Vishnu} M.,   {Vardhan} H.,  2023, \mn@doi [Astronomy \& Astrophysics]
  {10.1051/0004-6361/202243928}, \href
  {https://ui.adsabs.harvard.edu/abs/2023A&A...670A..54D} {670, A54}

\bibitem[\protect\citeauthoryear{Donoso-Oliva, Becker, Protopapas,
  Cabrera-Vives, Cádiz-Leyton  \& Moreno-Cartagena}{Donoso-Oliva
  et~al.}{2025}]{Donoso-Oliva_2025}
Donoso-Oliva C.,  Becker I.,  Protopapas P.,  Cabrera-Vives G.,  Cádiz-Leyton
  M.,   Moreno-Cartagena D.,  2025, arXiv preprint arXiv:2502.02717

\bibitem[\protect\citeauthoryear{Durkan, Bekasov, Murray  \&
  Papamakarios}{Durkan et~al.}{2020}]{nflows}
Durkan C.,  Bekasov A.,  Murray I.,   Papamakarios G.,  2020, {nflows}:
  normalizing flows in {PyTorch}, \mn@doi{10.5281/zenodo.4296287}, \url
  {https://doi.org/10.5281/zenodo.4296287}

\bibitem[\protect\citeauthoryear{Erhan, Bengio, Courville, Manzagol, Vincent
  \& Bengio}{Erhan et~al.}{2010}]{why_pretrain_dl_2010_erhan_bengio}
Erhan D.,  Bengio Y.,  Courville A.,  Manzagol P.-A.,  Vincent P.,   Bengio S.,
   2010, J. Mach. Learn. Res., 11, 625–660

\bibitem[\protect\citeauthoryear{Fraga et~al.,}{Fraga
  et~al.}{2024}]{Fraga_2024}
Fraga B. M.~O.,  et~al., 2024, arXiv preprint arXiv:2404.08798v2

\bibitem[\protect\citeauthoryear{Gagliano, Contardo, Foreman-Mackey, Malz  \&
  Aleo}{Gagliano et~al.}{2023}]{Gagliano_2023}
Gagliano A.,  Contardo G.,  Foreman-Mackey D.,  Malz A.~I.,   Aleo P.~D.,
  2023, \mn@doi [The Astrophysical Journal] {10.3847/1538-4357/ace326}, 954, 6

\bibitem[\protect\citeauthoryear{{Gaia Collaboration} et~al.,}{{Gaia
  Collaboration} et~al.}{2018}]{Babusiaux2018_ObservationalHRD}
{Gaia Collaboration} et~al., 2018, \mn@doi [A\&A]
  {10.1051/0004-6361/201832843}, 616, A10

\bibitem[\protect\citeauthoryear{Goyal et~al.,}{Goyal
  et~al.}{2021}]{Goyal_2021}
Goyal P.,  et~al., 2021, arXiv preprint arXiv:2103.01988v2

\bibitem[\protect\citeauthoryear{Guo, Pleiss, Sun  \& Weinberger}{Guo
  et~al.}{2017}]{Guo_2017}
Guo C.,  Pleiss G.,  Sun Y.,   Weinberger K.~Q.,  2017, arXiv preprint
  arXiv:1706.04599v2

\bibitem[\protect\citeauthoryear{Gupta, Muthukrishna  \& Audenaert}{Gupta
  et~al.}{2025b}]{Gupta_2025_SBPT}
Gupta R.,  Muthukrishna D.,   Audenaert J.,  2025b, arXiv preprint
  arXiv:2510.12958

\bibitem[\protect\citeauthoryear{Gupta, Muthukrishna, Rehemtulla  \&
  Shah}{Gupta et~al.}{2025a}]{Gupta_2025}
Gupta R.,  Muthukrishna D.,  Rehemtulla N.,   Shah V.,  2025a, arXiv preprint
  arXiv:2502.18558v2

\bibitem[\protect\citeauthoryear{Gómez, Neira, Hoyos, Arbeláez  \&
  Forero-Romero}{Gómez et~al.}{2020}]{Gómez_2020}
Gómez C.,  Neira M.,  Hoyos M.~H.,  Arbeláez P.,   Forero-Romero J.~E.,
  2020, arXiv preprint arXiv:2004.13877v2

\bibitem[\protect\citeauthoryear{Hinton, Vinyals  \& Dean}{Hinton
  et~al.}{2015}]{Hinton_2015}
Hinton G.,  Vinyals O.,   Dean J.,  2015, arXiv preprint arXiv:1503.02531

\bibitem[\protect\citeauthoryear{Ishida et~al.,}{Ishida
  et~al.}{2019}]{Ishida_2019}
Ishida E. E.~O.,  et~al., 2019, arXiv preprint arXiv:1909.13260v2

\bibitem[\protect\citeauthoryear{{Jamal} \& {Bloom}}{{Jamal} \&
  {Bloom}}{2020}]{Jamal_2020}
{Jamal} S.,  {Bloom} J.~S.,  2020, \mn@doi [Astrophysical Journal Supplement
  Series] {10.3847/1538-4365/aba8ff}, \href
  {https://ui.adsabs.harvard.edu/abs/2020ApJS..250...30J} {250, 30}

\bibitem[\protect\citeauthoryear{Koh et~al.,}{Koh et~al.}{2020}]{Koh_2020}
Koh P.~W.,  et~al., 2020, arXiv preprint arXiv:2012.07421v3

\bibitem[\protect\citeauthoryear{{LSST Science Collaborations}}{{LSST Science
  Collaborations}}{2009}]{lsstsciencebook2009}
{LSST Science Collaborations} 2009, The LSST Science Book.
LSST Corporation, \url {http://www.lsst.org}

\bibitem[\protect\citeauthoryear{Lee, Yang, von Davier, Forlizzi  \& Das}{Lee
  et~al.}{2023}]{Lee_2023}
Lee H.-P.,  Yang Y.-J.,  von Davier T.~S.,  Forlizzi J.,   Das S.,  2023, arXiv
  preprint arXiv:2310.07879v2

\bibitem[\protect\citeauthoryear{Li, Baptista  \& Marzouk}{Li
  et~al.}{2024}]{Li_2024}
Li F.,  Baptista R.,   Marzouk Y.,  2024, arXiv preprint arXiv:2411.08390v2

\bibitem[\protect\citeauthoryear{Li, Chen, Lin, Rehemtulla, Shah, Wu, Miller
  \& Liu}{Li et~al.}{2025}]{Li_2025}
Li W.,  Chen H.-Y.,  Lin Q.,  Rehemtulla N.,  Shah V.~G.,  Wu D.,  Miller
  A.~A.,   Liu H.,  2025, arXiv preprint arXiv:2510.06200

\bibitem[\protect\citeauthoryear{Lin, Goyal, Girshick, He  \& Dollár}{Lin
  et~al.}{2017}]{Lin_2017}
Lin T.-Y.,  Goyal P.,  Girshick R.,  He K.,   Dollár P.,  2017, arXiv preprint
  arXiv:1708.02002v2

\bibitem[\protect\citeauthoryear{Liu, Lin, Cao, Hu, Wei, Zhang, Lin  \&
  Guo}{Liu et~al.}{2021}]{Liu_2021}
Liu Z.,  Lin Y.,  Cao Y.,  Hu H.,  Wei Y.,  Zhang Z.,  Lin S.,   Guo B.,  2021,
  arXiv preprint arXiv:2103.14030v2

\bibitem[\protect\citeauthoryear{Lochner, McEwen, Peiris, Lahav  \&
  Winter}{Lochner et~al.}{2016}]{Lochner_2016}
Lochner M.,  McEwen J.~D.,  Peiris H.~V.,  Lahav O.,   Winter M.~K.,  2016,
  \mn@doi [The Astrophysical Journal Supplement Series]
  {10.3847/0067-0049/225/2/31}, 225, 31

\bibitem[\protect\citeauthoryear{{Long}, {El Karoui}, {Rice}, {Richards}  \&
  {Bloom}}{{Long} et~al.}{2012}]{Long_2012}
{Long} J.~P.,  {El Karoui} N.,  {Rice} J.~A.,  {Richards} J.~W.,   {Bloom}
  J.~S.,  2012, \mn@doi [Publications of the Astronomical Society of the
  Pacific] {10.1086/664960}, \href
  {https://ui.adsabs.harvard.edu/abs/2012PASP..124..280L} {124, 280}

\bibitem[\protect\citeauthoryear{{Malanchev}}{{Malanchev}}{2023}]{elasticc_dataset_abstract_only}
{Malanchev} K.,  2023, in American Astronomical Society Meeting Abstracts
  \#241. p. 117.03

\bibitem[\protect\citeauthoryear{Manning, Raghavan  \& Schütze}{Manning
  et~al.}{2008}]{manning2008introduction}
Manning C.~D.,  Raghavan P.,   Schütze H.,  2008, An Introduction to
  Information Retrieval.
Cambridge University Press, Cambridge, England, \url
  {https://www-nlp.stanford.edu/IR-book/}

\bibitem[\protect\citeauthoryear{Mohri \& Medina}{Mohri \&
  Medina}{2012}]{Mohri_2012}
Mohri M.,  Medina A.~M.,  2012, arXiv preprint arXiv:1205.4343v2

\bibitem[\protect\citeauthoryear{Moreno-Cartagena, Protopapas, Cabrera-Vives,
  Cádiz-Leyton, Becker  \& Donoso-Oliva}{Moreno-Cartagena
  et~al.}{2025}]{morenocartagena2025leveragingpretrainedvisualtransformers}
Moreno-Cartagena D.,  Protopapas P.,  Cabrera-Vives G.,  Cádiz-Leyton M.,
  Becker I.,   Donoso-Oliva C.,  2025, Leveraging Pre-Trained Visual
  Transformers for Multi-Band Photometric Light Curve Classification
  (\mn@eprint {arXiv} {2502.20479}), \url {https://arxiv.org/abs/2502.20479}

\bibitem[\protect\citeauthoryear{Muthukrishna, Narayan, Mandel, Biswas  \&
  Hložek}{Muthukrishna et~al.}{2019}]{Muthukrishna_2019}
Muthukrishna D.,  Narayan G.,  Mandel K.~S.,  Biswas R.,   Hložek R.,  2019,
  arXiv preprint arXiv:1904.00014v2

\bibitem[\protect\citeauthoryear{Möller \& de Boissière}{Möller \&
  de Boissière}{2019}]{Möller_2019_SuperNNova}
Möller A.,  de Boissière T.,  2019, \mn@doi [Monthly Notices of the Royal
  Astronomical Society] {10.1093/mnras/stz3312}, 491, 4277

\bibitem[\protect\citeauthoryear{{Naul}, {Bloom}, {P{\'e}rez}  \& {van der
  Walt}}{{Naul} et~al.}{2018}]{Naul_2018}
{Naul} B.,  {Bloom} J.~S.,  {P{\'e}rez} F.,   {van der Walt} S.,  2018, \mn@doi
  [Nature Astronomy] {10.1038/s41550-017-0321-z}, \href
  {https://ui.adsabs.harvard.edu/abs/2018NatAs...2..151N} {2, 151}

\bibitem[\protect\citeauthoryear{Opitz \& Burst}{Opitz \&
  Burst}{2019}]{Opitz_2019}
Opitz J.,  Burst S.,  2019, arXiv preprint arXiv:1911.03347v3

\bibitem[\protect\citeauthoryear{Perrigo}{Perrigo}{2023}]{perrigo2023openai}
Perrigo B.,  2023, Exclusive: {OpenAI} Used {Kenyan} Workers on {Less} Than
  {\$2} Per Hour to Make {ChatGPT} Less Toxic, TIME Magazine, \url
  {https://time.com/6247678/openai-chatgpt-kenya-workers/}

\bibitem[\protect\citeauthoryear{{Pimentel}, {Est{\'e}vez}  \&
  {F{\"o}rster}}{{Pimentel} et~al.}{2023}]{Pimentel_2023}
{Pimentel} {\'O}.,  {Est{\'e}vez} P.~A.,   {F{\"o}rster} F.,  2023, \mn@doi
  [Astronomical Journal] {10.3847/1538-3881/ac9ab4}, \href
  {https://ui.adsabs.harvard.edu/abs/2023AJ....165...18P} {165, 18}

\bibitem[\protect\citeauthoryear{Qu, Sako, Möller  \& Doux}{Qu
  et~al.}{2021}]{Qu_2021}
Qu H.,  Sako M.,  Möller A.,   Doux C.,  2021, arXiv preprint arXiv:2106.04370

\bibitem[\protect\citeauthoryear{Radford et~al.,}{Radford
  et~al.}{2021}]{Radford_2021_CLIP}
Radford A.,  et~al., 2021, CoRR, abs/2103.00020

\bibitem[\protect\citeauthoryear{Raji, Kumar, Horowitz  \& Selbst}{Raji
  et~al.}{2022}]{Raji_2022}
Raji I.~D.,  Kumar I.~E.,  Horowitz A.,   Selbst A.~D.,  2022, arXiv preprint
  arXiv:2206.09511v2

\bibitem[\protect\citeauthoryear{Rehemtulla et~al.,}{Rehemtulla
  et~al.}{2024}]{Rehemtulla_2024}
Rehemtulla N.,  et~al., 2024, arXiv preprint arXiv:2401.15167

\bibitem[\protect\citeauthoryear{Richards et~al.,}{Richards
  et~al.}{2011}]{Richards_2012}
Richards J.~W.,  et~al., 2011, \mn@doi [The Astrophysical Journal]
  {10.1088/0004-637X/744/2/192}, 744, 192

\bibitem[\protect\citeauthoryear{Rizhko \& Bloom}{Rizhko \&
  Bloom}{2024}]{Rizhko_2024}
Rizhko M.,  Bloom J.~S.,  2024, arXiv preprint arXiv:2411.08842

\bibitem[\protect\citeauthoryear{Shah, Gagliano, Malanchev, Narayan  \&
  Collaboration}{Shah et~al.}{2025}]{Shah_2025}
Shah V.~G.,  Gagliano A.,  Malanchev K.,  Narayan G.,   Collaboration T. L. D.
  E.~S.,  2025, arXiv preprint arXiv:2501.01496

\bibitem[\protect\citeauthoryear{Su, Lu, Pan, Murtadha, Wen  \& Liu}{Su
  et~al.}{2021}]{Su_2021}
Su J.,  Lu Y.,  Pan S.,  Murtadha A.,  Wen B.,   Liu Y.,  2021, arXiv preprint
  arXiv:2104.09864v5

\bibitem[\protect\citeauthoryear{Szegedy, Vanhoucke, Ioffe, Shlens  \&
  Wojna}{Szegedy et~al.}{2015}]{Szegedy_2015}
Szegedy C.,  Vanhoucke V.,  Ioffe S.,  Shlens J.,   Wojna Z.,  2015, arXiv
  preprint arXiv:1512.00567v3

\bibitem[\protect\citeauthoryear{Tan, Protopapas, Cádiz-Leyton, Cabrera-Vives,
  Donoso-Oliva  \& Becker}{Tan et~al.}{2025}]{Tan_2025}
Tan A.,  Protopapas P.,  Cádiz-Leyton M.,  Cabrera-Vives G.,  Donoso-Oliva C.,
    Becker I.,  2025, arXiv preprint arXiv:2509.24134

\bibitem[\protect\citeauthoryear{Vaswani, Shazeer, Parmar, Uszkoreit, Jones,
  Gomez, Kaiser  \& Polosukhin}{Vaswani et~al.}{2017}]{Vaswani_2017}
Vaswani A.,  Shazeer N.,  Parmar N.,  Uszkoreit J.,  Jones L.,  Gomez A.~N.,
  Kaiser L.,   Polosukhin I.,  2017, arXiv preprint arXiv:1706.03762v4

\bibitem[\protect\citeauthoryear{Vidal, Gagliano  \& Cuesta-Lazaro}{Vidal
  et~al.}{2025}]{Vidal_2025}
Vidal E.~P.,  Gagliano A.~T.,   Cuesta-Lazaro C.,  2025, arXiv preprint
  arXiv:2510.14202

\bibitem[\protect\citeauthoryear{Villar et~al.,}{Villar
  et~al.}{2019}]{Villar_2019}
Villar V.~A.,  et~al., 2019, \mn@doi [The Astrophysical Journal]
  {10.3847/1538-4357/ab418c}, 884, 83

\bibitem[\protect\citeauthoryear{Woo, Liu, Kumar, Xiong, Savarese  \&
  Sahoo}{Woo et~al.}{2024}]{Woo_2024}
Woo G.,  Liu C.,  Kumar A.,  Xiong C.,  Savarese S.,   Sahoo D.,  2024, arXiv
  preprint arXiv:2402.02592v2

\bibitem[\protect\citeauthoryear{Zhang, Helfer, Gagliano, Mishra-Sharma  \&
  Villar}{Zhang et~al.}{2024}]{Zhang_2024}
Zhang G.,  Helfer T.,  Gagliano A.~T.,  Mishra-Sharma S.,   Villar V.~A.,
  2024, arXiv preprint arXiv:2408.16829

\bibitem[\protect\citeauthoryear{Željko Ivezić et~al.,}{Željko Ivezić
  et~al.}{2008}]{Ivezić_2008}
Željko Ivezić et~al., 2008, arXiv preprint arXiv:0805.2366v5

\makeatother
\end{thebibliography}
